\documentclass[fleqn,usenatbib]{mnras}
\usepackage{newtxtext,newtxmath}
\usepackage{booktabs, siunitx, multirow, tabularx, adjustbox}
\usepackage[T1]{fontenc}

\DeclareRobustCommand{\VAN}[3]{#2}
\let\VANthebibliography\thebibliography
\def\thebibliography{\DeclareRobustCommand{\VAN}[3]
{##3}\VANthebibliography}

\usepackage{newtxtext,newtxmath}

\usepackage{graphicx}	
\usepackage{amsmath}	
\usepackage{pdflscape}
\usepackage{longtable}

\usepackage[encapsulated]{CJK}
\usepackage{ucs}
\usepackage[utf8x]{inputenc}
\newcommand{\tctext}[1]{\begin{CJK}{UTF8}{bkai}#1\ignorespacesafterend\end{CJK}}

\title[Formation and composition of the T1 planets]{Composition constraints of the TRAPPIST-1 planets from their formation}

\author[Childs et al.]{
Anna C. Childs$^{1,2,3}$ \thanks{E-mail: anna.childs@northwestern.edu}, Cody Shakespeare$^{2,3}$, David R. Rice$^{2,3}$, Chao-Chin Yang (\tctext{楊朝欽})$^{2,3,4}$,
\newauthor
 \, and Jason H. Steffen$^{2,3}$
\\
$^{1}$Center for Interdisciplinary Exploration and Research in Astrophysics (CIERA) and Department of Physics and Astronomy, Northwestern University, \\
1800 Sherman Ave, Evanston, IL 60201 USA\\
$^{2}$Nevada Center for Astrophysics, University of Nevada, Las Vegas,  Las Vegas, NV 89154, USA\\
$^{3}$Department of Physics and Astronomy, University of Nevada, Las Vegas, 4505 South Maryland Parkway, Las Vegas, NV 89154, USA\\
$^4$Department of Physics and Astronomy, The University of Alabama, Box~870324, Tuscaloosa, AL~35487-0324, USA
}
\date{Accepted XXX. Received YYY; in original form ZZZ}

\pubyear{2015}

\begin{document}
\label{firstpage}
\pagerange{\pageref{firstpage}--\pageref{lastpage}}
\maketitle

\begin{abstract}
We study the formation of the TRAPPIST-1 (T1) planets starting shortly after Moon-sized bodies form just exterior to the ice line.  Our model includes mass growth from pebble accretion and mergers, fragmentation, type-I migration, and eccentricity and inclination dampening from gas drag.  We follow the composition evolution of the planets fed by a dust condensation code that tracks how various dust species condense out of the disc as it cools.  We use the final planet compositions to calculate the resulting radii of the planets using a new planet interior structure code and explore various interior structure models.  Our model reproduces the broader architecture of the T1 system and constrains the initial water mass fraction of the early embryos and the final relative abundances of the major refractory elements.  We find that the inner two planets likely experienced giant impacts and fragments from collisions between planetary embryos often seed the small planets that subsequently grow through pebble accretion.  Using our composition constraints we find solutions for a two-layer model, a planet comprised of only a core and mantle, that match observed bulk densities for the two inner planets b and c.  This, along with the high number of giant impacts the inner planets experienced, is consistent with recent observations that these planets are likely dessicated.  However, two-layer models seem unlikely for most of the remaining outer planets which suggests that these planets have a primordial hydrosphere.  Our composition constraints also indicate that no planets are consistent with a core-free interior structure.
\end{abstract}




\section{Introduction}
The TRAPPIST-1 (T1) system is a late-M dwarf star that hosts seven tightly packed, terrestrial planets \citep{Gillon2016, Gillon2017}. The unique and tightly constrained planet masses and orbital distribution of this system has implications for its formation history and as a result, this system has been widely studied.  Observations show that the two inner-most planets have the largest masses in the system while the mass of the outer planets increases with their orbital distance \citep{Agol_2021}.  This mass trend has been referred to as a \emph{reversed mass ranking} and is difficult to explain with current planet formation theories \citep{Ogihara2022}.  The planets are in a complex resonance chain where the outer four planets are in first-order mean-motion resonances with each adjacent planet and the inner three planets are in higher-order resonances (8:5 and 5:3).  Three-body Laplace resonances also exist throughout the system \citep{Luger2017}.  

Terrestrial planets can primarily grow their mass though core accretion \citep{KI12} or through pebble accretion \citep[and references therein]{Johansen2017}.  \cite{Coleman2019} numerically modeled both mechanisms around a T1-like star to understand if planetary systems around low-mass stars preferentially form through either planetesimal or pebble accretion.  They found that while both planetesimal and pebble accretion form similar planetary systems, planets that formed through planetesimal accretion had a much larger water content than the planet analogues that formed through pebble accretion.  Thus, constraints on the water mass fraction (WMF) and bulk densities of the T1 planets can provide insight into their formation history.

Measurements from transit-timing variations and dynamical modeling helped constraining the bulk density of the planets \citep{Grimm2018,Dorn2018,Unterborn2018,Agol_2021}.  These studies showed that the planets have similar densities and are consistent with rocky worlds with water mass fractions (WMF) $<20\%$, suggesting that all the planets formed in a similar manner and their primary growth mode was via pebble accreiton.

\cite{Ormel2017} proposed that the formation of the T1 system first took place at the ice line where planetesimals formed by the streaming instability.  The streaming instability is when solid particles concentrate into dense filaments and undergo gravitational collapse into numerous bound objects.  Planetesimals up to $\sim$100\,km may form in this way \citep{JM15,Simon_2016,Schafer2017,LYS19}.  After the initial planetesimals form, they continue to grow by pebble accretion \citep{Johansen2017}.  Each planet then undergoes inward type-I migration and accretes silicate pebbles once it is inside the ice line.  In this manner, the planets form sequentially at the ice line.  The innermost planets stall near the inner edge of the gas disc, which is set by the star's magnetosphere.

The \cite{Ormel2017} formation theory has been the most widely accepted formation theory for the T1 system and as a result, it has been extensively tested. \cite{Lin2021} analytically tested this theory by evolving protoplanets in a gas disc that begin at the ice line and grow their mass via pebble accretion while migrating in.  Protoplanets sequentially appear every protoplanetary appearance timescale of $\sim 10^5$ orbital periods at the ice line.  This approach was able to reproduce the multiplicity and resonance structure of the T1 system.  However, analytical approaches neglect important effects such as two body interactions and the study was not able to reproduce the mass distribution of the T1 system.

\cite{Huang2021} tested the plausibility of the T1 planets forming sequentially at the ice line by numerically modeling the inward migration of the fully formed T1 planets from the ice line.  Using the $n$-body code \textsc{rebound} \citep{Rein2012}, and \textsc{reboundX} \citep{Tamayo2020} to model the effects of an evolving gas disc on the planets. \cite{Huang2021} demonstrated that if the T1 planets were sequentially produced and migrated inwards, the planets naturally converged into a chain of first order resonances.  Modeling a migration barrier in the inner gas-free cavity, where planets were pushed further inwards from an outer Lindblad torque from the gas disc, tidal forces, and orbital damping from gas drag, they were able to reproduce the observed two-body and three-body resonances of the system.  However, this mechanism does not explain the composition of the planets, since beginning with fully formed planets at the ice line may imply a water content much larger than what is observed.

\cite{Schoonenberg_2019} numerically modeled the formation of the T1 system starting with results from a Lagrangian dust evolution code that modeled the formation of planetesimals via the streaming instability \citep{Liu2019}.  They tracked the growth of planetesimals into planets via pebble accretion and planetesimal accretion using a modified version of the $n$-body code \textsc{mercury} \citep{Chambers1999} which included pebble accretion and the gas effects of type-I migration and aerodynamic drag.  Once the planets migrated to shorter orbits, they modeled the final stage of mass evolution semi-analytically in order to reduce computation time.  This approach reproduced the general mass distribution of the T1 planets, but not the orbital architecture as the last stages were not modeled numerically.

\cite{Ogihara2022} also successfully reproduced the masses of the T1 planets by numerically modeling the growth of embryos in a gas disc that loses mass from photoevaporation and disc winds in addition to gas accretion onto the central star.  The more complex temporal evolution of this gas disc results in an initially fast and then later, when the surface density profile flattens out, slow migration of the embryos.  They found that this fast-then-slow migration resulted in systems that display the reversed mass ranking mass trend found in T1.  However, their simulations began with relatively large embryos that already reached the pebble isolation mass, distributed between $0.015$--$0.2$\,au, after the disc has evolved for $1$\,Kyr. 

Although previous $n$-body studies have reproduced the broader features of the T1 system, it has proven difficult to reproduce the planet densities and masses, and the orbital architecture of the system when starting from an early stage in planet formation.  In this paper, we use a suite of numerical tools to constrain the bulk compositions of the planets by following their formation process starting just after Moon-sized bodies form and up until the gas disc dissipates, while reproducing the observed planet densities.  Detailed modeling of this period allows us to follow the composition of the planets throughout their formation process.  While cometary impactors can build or destroy the atmospheres of the T1 planets at a later time \citep{Kral2018}, and other late-stage mechanisms can alter the surface properties of the planets, we provide constraints on the bulk compositions of the T1 planets from their formation.

We improve upon previous work by using a disc model that changes in time and has not yet been used to study the T1 system, resolving solid body collisions in a more realistic manner (i.e. fragmentation), and by using the most up-to-date prescriptions for pebble accretion.  Furthermore, we provide the abundances of the refractory elements of the T1 planets for the first time and use these results to probe the interior structure of the planets using a new planetary interior structure code.

We present a new module for \textsc{reboundX} that tracks pebble accretion growth, type-I migration, and eccentricity and inclination damping from gas drag.  Our simulations also model the fragmentation of solid bodies involved in collisions \citep{ChildsSteffen2022}.  We place tight constraints on the composition of the bodies by simultaneously modeling the composition of the accreted dust pebbles as a function of location and time.  The composition of the dust is determined by a dust condensation code by \cite{Li2020} that tracks how dust condenses out of an evolving protoplanetary disc as it cools in time.  We use our final constraints on bulk composition to calculate the planet radius and density using the planetary interior structure code \textsc{Magrathea} \citep{Huang2022}.  

In Section \ref{sec:models} we describe our evolving gas disc and our prescriptions for pebble accretion, type-I migration, and eccentricity and inclination damping.  We also describe the model we use to track the composition evolution of the bodies.  In Section \ref{sec:n-body} we describe our $n$-body setup.  In Section \ref{sec:results} and Section \ref{sec:Composition} we lay out our results and in Section \ref{sec:discussion} we discuss the implications of these results and caveats of our models.  Lastly, we summarise our results in Section \ref{sec:conclusions}.

\section{Models} \label{sec:models}
In this section we discuss our models for disc evolution, mass growth via pebble accretion, and the effects of gas on the dynamics of the solid bodies.  We include gas effects throughout the duration of the simulation because the T1 planets are thought to have formed in less than a few Myr, a timescale shorter than the disc lifetime \citep{Raymond2021}.  However, the effects from the gas disc (i.e.\ pebble accretion, type-I migration, eccentricity and inclination damping) are turned off after a body moves inwards past $0.02 \, \rm au$ because the disc is thought to be truncated by the magnetosphere of the star \citep{Coleman2019}.  We implement these prescriptions into a new module for \textsc{reboundX} \citep{Tamayo2020} that works in tandem with the $n$-body integrator \textsc{rebound} \citep{Rein2012}.  We also describe how we track the evolution of the composition of the planets throughout the planet formation process, which is done as a post-processing step.

\begin{table}
\centering
\caption{The disc parameters adopted for our models.  We choose parameters that are consistent with a T-1 characteristic protoplanetary disc.}
 \vspace{1cm}
\begin{tabular}{ccc}
  \hline
    {Symbol} & {Parameter} & {Value}\\
    \hline   
{$M_{\star}$} & {stellar mass} & {$0.09 \, M_{\odot}$} \\

{$L_{\star}$} & {stellar luminosity} & {$1.5 \times 10^{-3} \, L_{\odot}$} \\

{$M_{\rm d}$} & {initial disc mass} & {$0.03 \, M_{\star}$} \\
{$f_{\rm p}$} & {scale factor for the pebble column density} & {1} \\
{$\Delta v$} & {sub-Keplerian speed of the gas} & {$29.9 \, \rm m \, s^{-1}$}\\

{$\alpha$} & {disc viscosity parameter} & {$10^{-4}$} \\
{$\tau_{\rm s}$} & {dimensionless stopping time of the pebbles} & {0.1 or 0.001}\\
{$r_{\rm ice}$} & {water ice line} & {$0.1 \, \rm au$} \\
{$\rho_{\rm pl}$} & {bulk density of planetesimals} & {$1.5 \, \rm g\, cm^{-3}$} \\
{$r_{\rm in}$} & {disc inner edge} & {$0.01 \, \rm au$} \\
{$R_1$} & {initial radius scale of the disc} & {$500 \, \rm au$} \\
  \hline
\end{tabular}
\label{tab:sim_parameters}
\end{table}

\subsection{Disc Evolution} \label{SS:disc}

Our disc evolution model is based on the accretion and evolution of T Tauri discs from \cite{Hartmann_1998}.  We adopt parameter values that are most fitting for the T1 system.  The surface density for a gas disc with mass $M_{\rm d}$ follows
\begin{equation}\label{eq:surface_profile}
    \Sigma(r,t_1) = \frac{M_{\rm d}}{2 \pi R_1^2} \frac{1}{(r/R_1)t_{1}^{3/2}} e^{-(r/R_1)/t_{1}}
\end{equation}
with orbital distance $r$ from the star and a radius scale of $R_1$.  $R_1$ is the radius within which $\sim$0.6 of the disc mass resides initially.  We set $R_1=500 \, \rm au$ and the initial disc mass $M_{\rm d}=0.03 M_{\star}$.  Observations of CO line emissions suggests gas discs are between 100-1000 au in size (see Figure 4 of \cite{Andrews2020}).  We choose an intermediate value that results in reasonable migration timescales of our starting solid bodies.  $t_1$ is a dimensionless time defined as 
\begin{equation}
    t_1=t/t_{\rm s}+1
\end{equation}
at time $t$.  The viscous timescale for the gas disc is
\begin{equation} \label{E:viscous_time}
    t_{\rm s}= \frac{1}{3}\frac{R_1^2}{\nu_1},
\end{equation}
where $\nu_1 = \nu(R_1)$ is the kinematic viscosity at $r = R_1$. 

Equation~\eqref{eq:surface_profile} assumes that the disc is vertically isothermal and has a radial temperature profile of $T(r) \propto r^{-1/2}$.  As in the minimum mass solar nebula \citep[MMSN;][]{Weidenschilling1977, Hayashi1981}, we adopt the temperature profile of
\begin{equation}\label{eq:Temp_profile}
    T(r)=280(r/ \rm au)^{-1/2}(L_{\star}/L_{\odot})^{1/4}\, \rm K
\end{equation}
where $L_{\star}$ is the luminosity of the star.  For consistency with previous studies of the T1 system, we place the ice line at $r_{\rm ice}=0.1 \, \rm au$.  Assuming the ice line corresponds to where the temperature is $170 \, \rm K$, Equation~\eqref{eq:Temp_profile} leads to a stellar luminosity of $L_{\star} \approx 1.5 \times 10^{-3} \, L_{\odot}$.  This luminosity is three times larger than the current luminosity of the T1 star, and therefore our model considers a much earlier time in the life of the star.  While the location of the ice line will change in time we opt to use a fixed location of the ice line for simplicity.

The viscosity of the gas disc is prescribed by $\nu=\alpha c_{\rm s} H$, where $\alpha$ is a dimensionless constant, $c_\mathrm{s}$ the speed of sound, and $H$ the vertical scale height of the gas \citep{Shakura1977}.
The speed of sound is given by
\begin{equation}
    c_{\rm s}=\sqrt{\frac{k_{\rm B} T(r)}{\mu m_{\rm H}}}
\end{equation}
where $k_{\rm B}$ is the Boltzmann constant, $\mu = 2.34$ is the mean molecular weight of the gas, and $m_\mathrm{H}$ is the hydrogen mass.  In turn, the vertical scale height of the gas is given by
\begin{equation}
    H = c_{\rm s}/\Omega,
\end{equation}
where $\Omega=\sqrt{G M_{\star}/r^3}$ is the Keplerian angular frequency.  We adopt $\alpha=1 \times 10^{-4}$, and along with $R_1$, Equation~\eqref{E:viscous_time} sets the timescale for the viscous evolution of the gas disc.

Finally, we assume that the column density of the pebbles is determined by a constant dust-to-gas ratio, that is,
\begin{equation}
\Sigma_{\rm p}(r)=0.01f_{\rm p} \Sigma(r)
\end{equation}
where $f_{\rm p}$ is a constant scale factor.  We adopt a dust-to-gas ratio of 1\% and hence $f_\mathrm{p} = 1$.

\subsection{Pebble accretion}
Mass growth of a planetesimal via pebble accretion is separated into either the Bondi regime or the Hill regime, depending on the size of the pebbles and the mass of the central planetesimal \citep{Johansen2017}.  Pebbles move from the Bondi regime to the Hill regime once the radius of the central planetesimal reaches the transition radius,
\begin{equation}
    R_{\rm t}(r)=1160{\rm km} \left ( \frac{r}{5 \, \rm au} \right)^{1/2} \left ( \frac{\Delta v}{30 {\,\rm m\, s^{-1}}} \right ) \left ( \frac{\rho_{\rm pl}}{{\rm2\times10^3 kg\, m^{-3}}} \right )^{-1/3}
\end{equation}
where $\rho_{\rm pl}$ is the bulk density of the planetesimal and $\Delta v$ can be approximated by the sub-Keplerian speed of the gas when the pebbles are at least marginally coupled to the gas \citep{Lambrechts2012}.  In our models, $\Delta v$ is then
\begin{equation}
    \Delta v = -\frac{1}{2}\frac{H}{r}\frac{\partial\ln P}{\partial\ln r}c_{\rm s} \approx \frac{11}{8}\frac{c_{\rm s}^2}{v_{\rm K}}=29.9\,\textrm{m\,s}^{-1},
\end{equation}
where $P$ is the pressure in the mid-plane of the disc and $v_{\rm K}= \sqrt{GM_{\star}/r}$ is the Keplerian velocity, and we have used in the last two steps the profiles in Section~\ref{SS:disc}.



In the 3D-Bondi branch, mass growth from pebble accretion proceeds as

\begin{equation}\label{eq:mass_growth}
\begin{split}
   &\Dot{M} = 8.4 \times 10^{-3} M_{\oplus} {\rm Myr^{-1}} f_{\rm p}
\left ( \frac{m_{\rm pl}}{10^{-4}M_{\oplus}} \right)^2 \left ( \frac{\Delta v}{30 {\rm m\,s^{-1}}} \right)^{-3} \left(\frac{H_{\rm p}/H}{0.1} \right)^{-1} \\
& \times \left(\frac{H/r}{0.05} \right)^{-1}\left (\frac{r}{5 \rm \, au} \right )^{-2}
   \end{split}
\end{equation}

Once the radius of the planetesimal reaches the transition radius, mass growth proceeds in the 2D-Hill branch as
\begin{equation}
    \Dot{M}=2 R_{\rm acc} \Sigma_{\rm p}\left ( \Delta v + \Omega R_{\rm acc}\right ),
\end{equation}
where $R_{\rm acc}$ is the radius from which the planetesimal accretes pebbles.
The accretion radius can be approximated by
\begin{equation}\label{eq:accretion_radius}
    R_{\rm acc} = \left(\frac{\Omega \tau_{\rm f}}{0.1} \right)^{1/3} R_{\rm H} \mathrm{exp}\left [-0.4 \left( \tau_{\rm f}/ t_{\rm p} \right )^{0.65} \right ],
\end{equation}
where $R_{\rm H}=(m_{\rm pl}/3M_{\star})^{1/3}r$ is the Hill raidus, $t_{\rm p}=Gm_{\rm pl}/(\Delta v + \Omega R_{\rm H})^3$ is the characteristic timescale for pebbles to pass by the planetesimal, and $\tau_{\rm f}$ is the friction time \citep{Ormel2010}.

Mass growth continues on the 2D-Hill branch until the planetesimal reaches the isolation mass when the planetesimal is large enough to induce a pressure bump exterior to its orbit which halts the incoming flux of pebbles.  We adopt the pebble isolation mass (PIM) from \cite{Ataiee2018},
\begin{equation}
    \mathrm{PIM} = (H/r)^3 \sqrt{37.3 \alpha + 0.01} \times \left [ 1 + 0.2 \left ( \frac{\sqrt{\alpha}}{H/r} \sqrt{\frac{1}{\tau_{\rm s}^2}+4} \right)^{0.7}\right ] M_{\star},
\end{equation}
where $\tau_\mathrm{s} \equiv \Omega\tau_\mathrm{f}$ is the dimensionless stopping time.  The dimensionless stopping time $\tau_\mathrm{s}$ of the pebbles depends on their location and composition \citep{Birnstiel2012, Birnsteil2016}.  We use $\tau_{\rm s}=0.1$ at or exterior to the ice line, approximating the radial drift barrier, and $\tau_{\rm s}=0.001$ interior to the ice line which roughly corresponds to the fragmentation barrier \citep{DD14,BM22}.  We do not model a change in the pebble surface density profile at the ice line.  The assumption of a constant pebble flux is not trivial as planetesimals that form near the ice line will reduce the pebble flux in the inner disk \citep{Drazkowska2017}.

Once a given body reaches the PIM, all other bodies interior to it also stop growing by pebble accretion.  When the bodies accrete pebbles we do not decrease the pebble density in the local region since the pebble accretion rate does not appear to be comparable, let alone exceed, the radial pebble flux often needed in modeling \citep{Lambrechts2019}.  On the other hand, we do track the total pebble mass.  All bodies stop growing by pebble accretion once the total dust mass has been reached.

\subsection{Type-I migration and gas drag}
Angular momentum exchange via spiral density waves cause the planetesimals to migrate inwards via type-I migration, and gas drag dampens the eccentricity and inclination of a planetesimal.  \cite{Cresswell2008} empirically derived expressions for the accelerations a body experiences in a gas disc, which can be implemented into $n$-body code.  The accelerations a body experiences from type-I migration, eccentricity damping, and inclination damping are
\begin{equation}
    \textbf{a}_{\rm m}=-\frac{\textbf{v}}{t_{\rm m}},
\end{equation}

\begin{equation}
    \textbf{a}_{\rm e}=-2\frac{(\textbf{v} \cdot \textbf{r})\textbf{r}}{r^2t_{\rm e}},
\end{equation}

\begin{equation}
    \textbf{a}_{\rm i}=-\frac{v_z}{t_{\rm i}}\textbf{k},
\end{equation}
respectively. $\textbf{k}$ is the unit vector in the $z$-direction and $\textbf{v}$ and $\textbf{r}$ are the velocity and position vectors of the body. The timescales associated with each of these accelerations are scaled by the damping timescale
\begin{equation}
    t_{\rm wave}= \frac{M_{\star}}{m_{\rm pl}} \frac{M_{\star}}{\Sigma(r)r^2}\left ( \frac{H}{r} \right)^4 \Omega^{-1},
\end{equation}
from \cite{Tanka2002}.  The eccentricity damping time is
\begin{equation}
\begin{split}
   & t_e=\frac{t_{\rm wave}}{0.780}\times\\
   & \left[1-0.14\left(\frac{e}{H/r} \right)^2 +  0.06\left(\frac{e}{H/r} \right)^3  +0.18\left(\frac{e}{H/r} \right) \left(\frac{i}{H/r} \right)^2 \right],
    \end{split}
\end{equation}
and the inclination damping time is
\begin{equation}
\begin{split}
   & t_{\rm i}=\frac{t_{\rm wave}}{0.544}\times\\
   & \left[1-0.30\left(\frac{i}{H/r} \right)^2 +  0.24\left(\frac{i}{H/r} \right)^3  +0.14\left(\frac{i}{H/r} \right) \left(\frac{e}{H/r} \right)^2 \right].
    \end{split}
\end{equation}
The type-I migration timescale is
\begin{equation}
\begin{split}
&t_{\rm m} = \frac{2 t_{\rm wave}}{2.7+1.1 \beta}\left ( \frac{H}{r} \right)^{-2} \\
& \left( P(e) + \frac{P(e)}{|P(e)|} \left[ 0.07\left( \frac{i}{H/r}\right)+0.085\left( \frac{i}{H/r}\right)^4-0.08\left( \frac{e}{H/r}\right) \left( \frac{i}{H/r}\right)^2\right] \right)
\end{split}
\end{equation}
where 
\begin{equation}
    P(e)=\frac{1+\left( \frac{e}{2.25H/r}\right)^{1.2} + \left( \frac{e}{2.84H/r}\right)^6}{1-\left( \frac{e}{2.02H/r}\right)^4},
\end{equation}
and $\Sigma(r) \propto r^{-\beta}$.  Following our surface density profile adopted in Section \ref{SS:disc}, we set $\beta=1$. 

Dynamical studies have shown that a rapid migration of the fully formed planets is needed to break out of various three-body mean motion resonances (MMRs) before arriving in their current resonant chain.  Because these fast migration rates are needed to reproduce the resonant structures an efficient stalling mechanisms may have been present in the inner region of the disc to prevent the planets from falling into the star.  Rapid migration timescales of the fully formed T1 planets can naturally explain first-order MMRs in the system \citep{Huang2021}, but the inner two planets are observed to be in higher order MMRs which indicates divergent migration in the inner disc.  \cite{Liu2017} demonstrated that divergent migration can happen close to the star from magnetospheric rebound.  \cite{Huang2021} were able to reproduce the T1 resonant structure by modeling a strong negative torque in the inner cavity, although this divergent torque was not physically motivated.  These studies considered the dynamics of the fully formed T1 planets, when less gas is present.  Migration timescales throughout the formation process are likely much shorter when more gas is present, but scattering and resonances may help reduce the migration rates throughout the formation process.  

To reproduce the proposed stalling mechanism thought to exist in the T1 system, we use the ``inner\_disc\_edge'' module in \textsc{reboundX} by Kajtazi et al.\ (in prep.)  This module applies an inner disc edge that functions as a planet trap by applying an opposite and roughly equal magnitude torque on the migrating body that enters the planet trap.  We do not allow bodies to migrate past the orbit of the innermost T1 planet, $\sim  0.01 \, \rm au$ by setting the inner disc edge to $0.01 \, \rm au$ and the width of the planet trap to $0.01 \, \rm au$.  The region in which this planet trap is employed is between $0.01-0.02 \, \rm au$.  All our parameter choices for our fiducial disc evolution model are listed in Table~\ref{tab:sim_parameters}.

\subsection{Composition evolution}
We use the dust condensation code by \cite{Li2020}, which models how dust condenses out of an evolving protoplanetary disc as the disc cools. The dust condensation code gives the initial elemental and mineral distributions of the protoplanetary disc that determines the composition of embryos that form at different orbital distances.  We then follow the composition evolution of the embryos as they collide with one another and grow via pebble accretion to form planets. The formation location and collision history of the embryos determines the resulting composition of the planets. The final Fe/Si molar ratio is used in \textsc{Magrathea} \citep{Huang2022} to determine the mass fraction of the planet's iron core and the planet's radius.

The dust condensation model is run independently of the evolution models discussed in previous sections. We use the dust condensation model for a solar-type star as presented in \cite{Li2020} as this is the system the code was developed for. We encountered difficulties converting the surface density profile to the one shown in Equation \ref{eq:surface_profile} and using disc parameters more compatible with an M-dwarf system.  We use the solar abundances in the condensation code since the measured metallicity of T1 is similar to the Sun's \citep{Gillon2016}.  For these reasons, we use the dust composition profile from a Sun-like star at a single epoch and re-scale the results to fit our T1 disc.  Successfully modeling dust condensation in an M-dwarf disc and comparing results with the re-scaled Sun-like disc, used here, will be the subject of future work.

Following the same solar model of \cite{Li2020}, the surface density profile for the dust condensation evolution is
\begin{equation}
    \Sigma (r,t)=\frac{M_{\rm disc}}{\pi r_0^2(t)}\exp\left\{-\left[\frac{r}{r_0(t)}\right]^2\right\},
    \label{eq:sigmadust}
\end{equation}
\noindent where $M_{\rm disc}=0.21M_\odot$ and $r_0(t)$ is the characteristic disc radius.  This disc mass corresponds to a disc around a Sun like star immediately after formation.  High accretion rates onto the star quickly deplete the disc mass and after one evolution timescale ($\sim 26 \, \rm Kyr$), the disc mass is less than 20\% its initial disc mass (see \cite{Cassen1996} Figure 2).
The temperature profile is
\begin{equation}
    T^4=\frac{3G\tau M_* \dot{M_I}}{64\pi \sigma_{SB}r^3},
    \label{eq:Tc}
\end{equation}
\noindent where $\dot{M_I}=$ is the accretion rate of gas, $\sigma_{SB}$ is the Stefan-Boltzmann constant, the optical depth is $\tau=\kappa \Sigma/2$, and $\kappa=4\text{ cm}^2\text{g}^{-1}$ is the opacity for a Solar nebula, the same used in \cite{Li2020}.  This temperature profile is for a disc dominated by viscous heating \cite{Cassen1996}.  The disc is not in a steady state and the mass accretion rate changes in time (see \cite{Li2020} Equation 8).
Finally, we re-scale the abundance distribution from the solar model to fit the size of our T1 disc by normalizing the locations of the ice line between the two discs.  The re-scaled dust distributions for 12 elements are shown in Figure \ref{fig:abundances}.  For more details on how the dust condensation results are re-scaled and the validity of extrapolating results from the solar system, see Appendix \ref{sec:appendix_B}.

The chemical equilibrium of condensing dust is modeled with GRAINS \citep{Petaev2009} which includes 33 different elements that form 520 condensed and 242 gaseous species. The combined disc evolution and chemical equilibrium code returns the relative abundance of the elements and condensed species as a function of orbital distance in the disc, at different times.  Further details on the dust evolution model can be found in \cite{Li2020}. 

Using this code, we determine the dust composition of our disc at any orbital distance to track the composition evolution of the solid bodies.  Our composition tracking code for the solid bodies is based on the composition tracking code of \cite{ChildsSteffen2022} and includes composition changes from pebble accretion.  The bodies all begin just exterior to the ice-line and we experiment with different refractory element:water-ice ratios.  The refractory elements for a given embryo are determined by the dust composition at the embryo's orbital distance, as determined by the condensation code.  

When two bodies collide, the target is the more massive one involved in the collision, with mass $M_{\rm t}$ and initial composition
\begin{equation}
    \textbf{X} = \left(x_1,x_2,...,x_{n}\right)
\end{equation}
where $x_i$ is the relative abundance of the $i^{\rm th}$ species such that
\begin{equation}
     \sum_{i=1}^{n} x_i =1.
\end{equation}
The projectile is the less massive involved in the collision, with mass $M_{\rm p}$ and initial composition 
\begin{equation}
    \textbf{Y} = \left(y_1,y_2,...,y_{n}\right)
\end{equation}
where $y_i$ is the relative abundance of the $i^{\rm th}$ species such that
\begin{equation}
     \sum_{i=1}^{n} y_i =1.
\end{equation}

If the collision results in an elastic bounce with no mass exchange, then the composition of each body remains the same.  If the collision results in a merger or partial accretion of the projectile, the composition of the target becomes
\begin{equation}
\mathbf{X'} = \frac{M_{\rm t}\mathbf{X} + M_{\rm p}'\mathbf{Y}}{M_{\rm t} + M_{\rm p}'}.
\end{equation}
where $M_{\rm p}'$ is the mass of the projectile that is accreted by the target.  If any fragments are produced from the projectile, they are assigned the composition of the projectile.

If the target becomes eroded, then the composition of the target remains the same but has a less mass $M_{\rm lr}$. The composition of the new fragment(s) becomes
\begin{equation}
\mathbf{X'} = \frac{M_{\rm diff}\mathbf{X} + M_{\rm p}\mathbf{Y}}{M_{\rm tot}-M_{\rm lr}}.
\end{equation}
where $M_{\rm tot}=M_{\rm t} + M_{\rm p}$ and $M_{\rm diff}=M_{\rm t} -M_{\rm lr}$.

We chronologically resolve all the collisions, using the prescription described above, and update the compositions of all the bodies according to the amount and location of the pebbles the body accumulated over the last $100 \, \rm years$.  The amount of pebbles a body has accumulated over $100 \, \rm years$, $M_{\rm peb}$, is the mass difference between the body at time $t$ and time $t+100 \, \rm years$, after all collisions have been accounted for.  The relative abundances of the pebbles is given by,
\begin{equation}
    \mathbf{Y}_{\rm peb} = \left(y_{\rm peb, 1},y_{\rm peb, 2},...,y_{\rm peb, \textit{n}}\right),
\end{equation}
where $y_{\rm peb, \textit{i}}$ is the relative abundance of the $i^{th}$ species for the pebble and
\begin{equation}
     \sum_{i=1}^{n} y_{{\rm peb},i} =1.
\end{equation}
$\mathbf{Y_{\rm peb}}$ is determined by the radial location of the pebble at time $t$ and the output from the dust condensation code.  The new composition of a body, which we refer to as the target, from pebble accretion is set by
\begin{equation}
    \mathbf{X'}= \frac{M_{\rm t}\mathbf{X} + M_{\rm peb}\mathbf{Y}_{\rm peb}}{M_{\rm t} + M_{\rm peb}}.
\end{equation}

\section{N-body setup}\label{sec:n-body}

In our $n$-body simulations we begin with 30 $0.01 \, \rm M_{\oplus}$ embryos exterior to the ice line.  The number of embryos are chosen so that the total initial embryo mass is $5\%$ of the mass of the T1 planetary system.  We distribute the embryos in an annulus just exterior to the ice line between $0.1-0.15 \, \rm au$.  The embryos have a different surface density than the gas profile as they preferentially form at the ice line and the formation of these large solid bodies leads to decoupling from the gas disc to some degree.  As a result, we choose for the embryos to follow a surface density profile of $\Sigma_{\rm pl} \propto r^{-3/2}$, which is a commonly used surface density profile for the starting bodies in studies of the solar system \citep{Weidenschilling1977}.  

We adopt a density of $1.5 \, \rm g \,cm^{-3}$ for our embryos which is consistent with 50\% ice and 50\% rock.  We also experiment with multiple initial WMFs to find WMFs that result in planet radii that match the observations.  We tested initial WMFs of 15\%, 20\%, and 25\% and find an initial WMF of 20\% for planetesimals resulted in planet radii that are in better agreement with the observed T1 planet radii (see Section \ref{sec:Composition}).  As a result, we report the results for two different initial compositions--the starting composition of the embryo is either 50\% water-ice and 50\% of the dust composition found at the embryo's initial radial location or, 20\% water-ice and 80\% of the dust composition found at the embryo's initial radial location.

All of the orbital elements for each body are chosen randomly and follow a uniform distribution.  The eccentricities ($e$) are distributed between $0.0-0.1$ and the inclinations ($i$) between $0.0^{\circ}-0.5^{\circ}$.  Because our model assumes planetesimals form in a narrow annulus just exterior to the ice line, motivated by results from \cite{Schafer2017} and \citep{Schoonenberg_2019} we use a relatively large value for eccentricity (see Section \ref{sec:ecc_inc} for more details on this point).  The longitude of ascending node ($\Omega$), argument of pericenter ($\omega$), and mean anomaly ($f$) are all distributed between $0^{\circ}-360^{\circ}$. 

Using the $n$-body code \textsc{rebound} and the \textsc{reboundX} module described above, we integrate 100 runs using the \textsc{mercurius} hybrid integrator.  We change the random seed generator in each run to provide variation of the orbital elements in the particle disc.  We integrate each run for $3 \, \rm Myr$.
Unless being inside the inner cavity which extends out to $0.01 \, \rm au$, all bodies experience growth via pebble accretion -- until they reach the PIM --, type-I migration, and eccentricity and inclination damping at all times.
Solid bodies are also free to interact with one another and collisions are resolved with fragmentation.  We set the minimum fragment mass to $0.01 \, M_{\oplus}$.  The fragmentation model we implement is detailed in \cite{ChildsSteffen2022}. A smaller fragment mass would be more realistic.  However, this mass is chosen as the result of computational limitations.

\section{System Architecture and Formation History}\label{sec:results}

Of the 100 runs we conducted, we first focus our analysis on the runs that returned systems with at least six planets since we are interested in systems similar to T1.  We define a planet as a body having a mass greater than or equal to $0.2 \, \rm M_{\oplus}$.  We find 24 runs that meet this criteria. Of these 24 runs, nine runs had six planets, two runs had seven planets, eight runs had eight planets, four runs had nine planets, and one run had 11 planets.  Interestingly, our model is four times more likely to produce eight planets instead of seven.  By analyzing the three-body resonance angles throughout the T1 system, the existence of an outer eighth planet in the system has been predicted \citep{Kipping2018}.  While \cite{Agol_2021} did not find evidence of TTVs from an eighth planet at a limited range of exterior orbital radii, our models produce systems with an eighth planet that has a mass similar to T1-h and is usually found exterior to the ice line.  

Table \ref{tab:system_data} lists the planet properties for each planet that formed in these 24 runs.  We list the simulation run number, the planet multiplicity (No.), and the mass ($M_{\rm p}$), semi-major axis ($a_{\rm p}$), eccentricity ($e$), inclination ($i$), the water mass fraction(WMF), iron (Fe), magnesium (Mg), silicon (Si), and oxygen (O) mass fractions (along with eight other elements) for each planet built from bodies that begin with a 20\% or 50\% WMF exclusively.  Lastly, we list the fraction of the final planet mass that came from pebble accretion (Peb), fragments (Frag), and embryo accretion (Em).

\subsection{Mass distributions}

\begin{figure*}
	\includegraphics[width=1.75\columnwidth,height=.9\textheight]{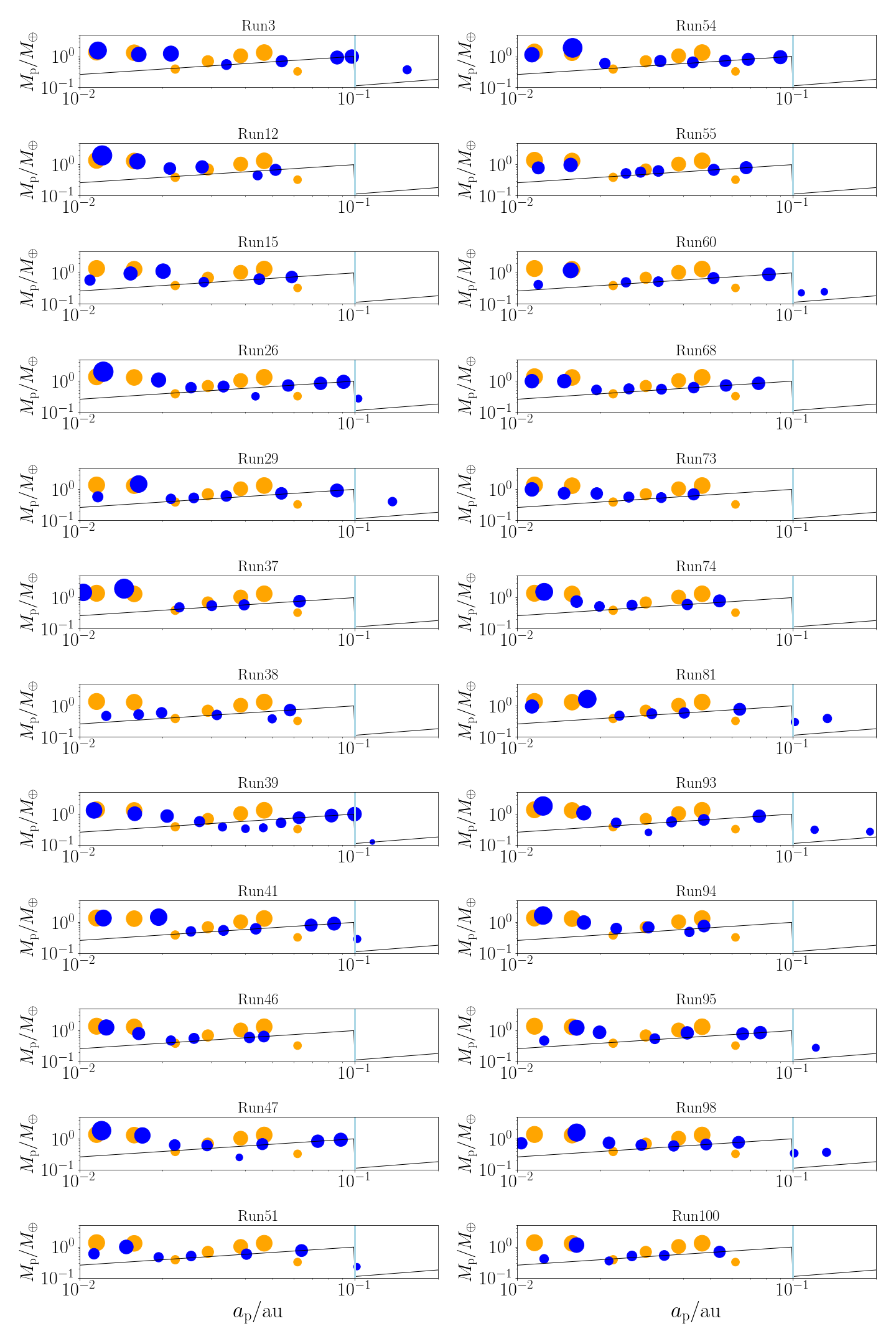}
    \caption{Mass ($M_{\rm p}$) vs. semi-major axis ($a_{\rm p}$) for all of the remaining planet-sized bodies in each run that contains at least six planets at time $t=3 \, \rm Myr$.  Our simulation data is shown in blue and the T1 planets are shown by orange points.  The point sizes are proportional to planet mass.  We also plot the location of the ice line by a blue vertical line and the PIM by a black line for reference.} 
    \label{fig:T1_runs}
\end{figure*}

Figure \ref{fig:T1_runs} shows the mass ($M_{\rm p}$) and semi-major axis ($a_{\rm p}$) of the planets that formed in the 24 runs.  The T1 planets are shown in orange and all the remaining bodies at $t=3 \, \rm Myr$ in a given run are shown in blue.  The size of the dots are proportional to the mass of the planet.  The PIM is marked by a black line and the ice line by a vertical blue line for reference.  The average total planetary mass of our simulated systems is $5.64 \, M_{\oplus}$ and the total mass of the T1 system is $6.45 \, M_{\oplus}$.  On average, each run grows its total initial embryo mass of $0.3 \, M_\oplus$ by almost 19 times ($\sim 5.64 \, M_{\oplus}$) through pebble accretion.

In the T1 system, the inner two planets are the most massive and then a reversed mass ranking is found for planets d-g where planet mass increases with semi-major axis. In 13 of the 24 runs, the inner most planets are the most massive planets in the system.  The bodies that first undergo runaway accretion and accrete the most embryos at the start of the simulation are the first to migrate inwards.  As the mass of the body increases, so does its migration rate and the body migrates inwards until it either reaches the inner cavity or is trapped by resonances with inner planets.  Since there are no resonances to halt the inward migration of the first protoplanets, this results in a build up of material at the inner edge. This build up of material at the inner edge leads to collisions and accretion that eventually builds the more massive inner planets.

The innermost planets typically accrete more embryos which allows them to grow more massive than the PIM in the inner disc region.  This finding is in agreement with \cite{Ogihara2022} although \cite{Ogihara2022} explored a formation pathway more akin to in-situ planet formation whereas we start with smaller bodies only exterior to the ice line. The bodies that avoid mergers at the start of the simulation migrate inwards at a later time when there are less planetesimals and embryos available to accrete and thus, grow majority of their mass via pebble accretion.  These subsequent planets that form grow quickly to the PIM once they cross the ice line.  As a result, most of the outer planet masses are near the PIM which increases with distance up to the ice line and can explain the reversed mass ranking of the T1 d-g planets.

The small size of the outermost planet is achieved when the planet forms exterior to the ice line, where the PIM is much lower due to a larger value for $\tau_{\rm s}$.  While this formation mechanism may explain the small size of T1-h, it is not clear if a planet that grows most of its mass exterior to the ice line is consistent with the observed density of T1-h.  In 10 of the 24 runs, the outermost planet is the smallest (or close to the smallest) but is found exterior to the ice line. This places our T1-h analogues at larger semi-major axes and results in planets with larger amounts of water than what is expected from the observed bulk density of T1-h (see Section~\ref{sec:Composition}).

\subsection{Period distributions}
\begin{figure}
	\includegraphics[width=1\columnwidth]{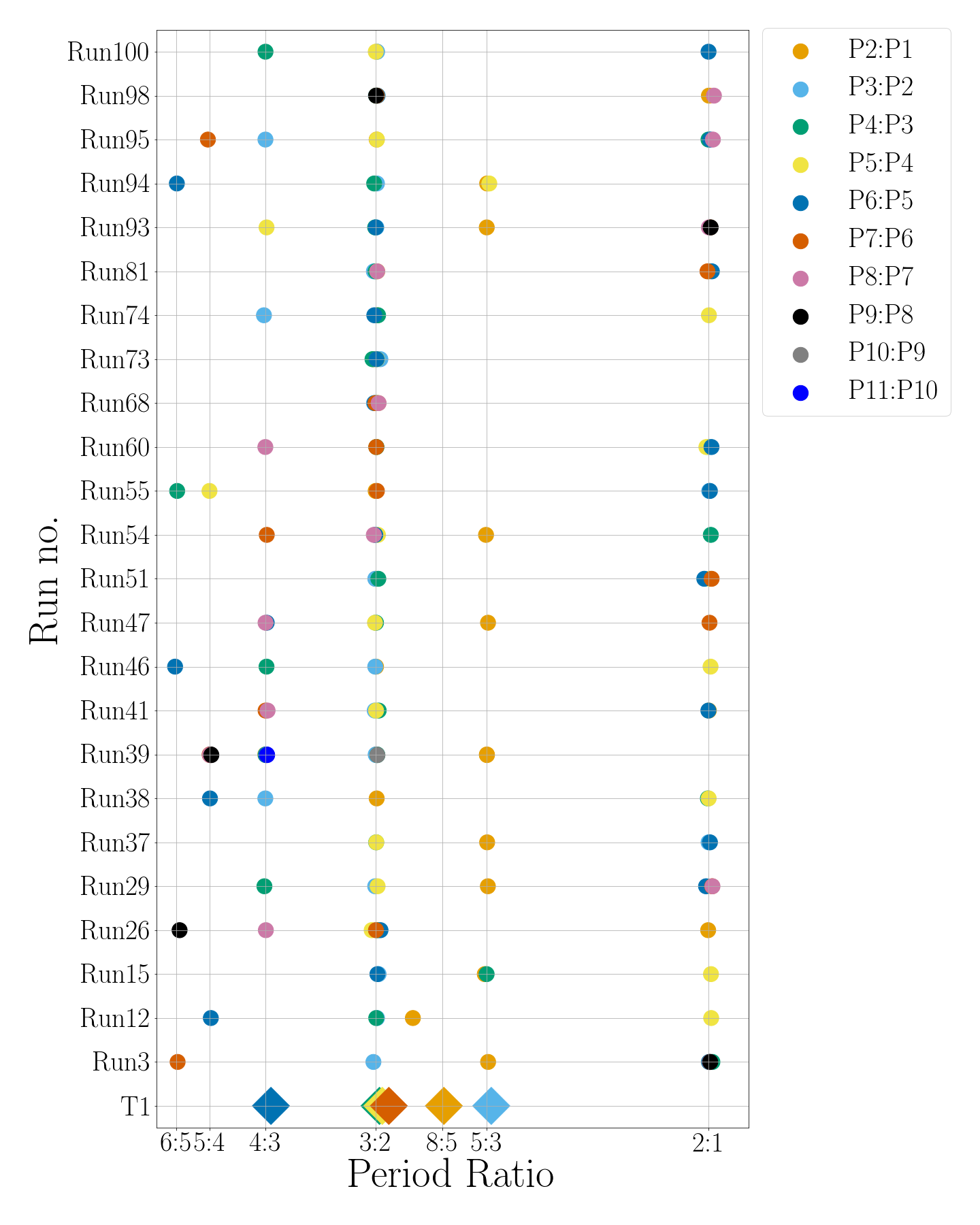}
    \caption{Period ratios of adjacent planets in all of our runs that produced six or more planets. Each planet pair is given a unique color.  T1 planet MMRs are denoted by diamonds.  Three T1 planet pairs are found near the 3:2 MMR.} 
    \label{fig:alpha_MMRs}
\end{figure}

Figure \ref{fig:alpha_MMRs} shows the period ratios found between adjacent planets in each of our 24 runs and also the period ratios of the T1 planets. Of the 24 runs, where we focus our analysis, the first-order 3:2 MMR is found in all 24 runs, 13 of the runs contain the first order 4:3 MMR and nine of the runs contain the second order 5:3 MMR.   While the stronger resonances of the T-1 system may be found in most of our runs we do not find the 8:5 MMR of the innermost planets observed in the T1 system in any of our 24 runs.  We attribute this to the simplified treatment of the inner disc cavity and lack of tidal effects.  However, close to an 8:5 MMR, a 2:1 MMR is found in all but four of the runs and some planets may also be found in 5:4 and 6:5 MMRs.  

\cite{Huang2021} demonstrated that once a more accurate treatment of the inner disc region is modeled, by incorporating the effects of an expanding gas-free cavity and the dynamics of the planets in this cavity, the fully formed inner two planets may break out of first order resonances and migrate into the observed 8:5 and 5:3 resonances.  Similarly, \cite{Ogihara2022} better recovered the observed resonances of the T1 system due to their use of a more complex disc evolution which resulted in more dynamic migration rates of the planets.

Three-body Laplace resonances may also be found throughout the T1 system.  These resonances contribute to the long term stability of the system \citep{Luger2017}.  The generalized three-body Laplace relation (GLR) angle is given by,
\begin{equation}
    \phi_{i,i+1,i+2}=p \lambda_i - (p+q) \lambda_{i+1}+q \lambda_{i+2},
\end{equation}
where $\lambda_i$ is the mean longitude of the $i$th planet, and $p$ and $q$ are integers.  The GLR is considered stable if the angle $\phi$ librates about $180^{\circ}$.  We consider the five main GLR angles observed in the T1 system (see \cite{Huang2021} for a review of these angles) in all of our runs that contain at least seven planets.  We do not find any of these five GLR resonances over the last 1 Myr of simulation time in any of the runs.  Again, this could be attributed to the assumed evolution of the gas disc \citep{Ogihara2022}, the assumed evolution of the disc's inner cavity \citep{Huang2022}, and/or the lack of tidal effects \citep{Papaloizou2018}.

\subsection{Eccentricity and inclination distributions}\label{sec:ecc_inc}
\begin{figure}
	\includegraphics[width=\columnwidth]{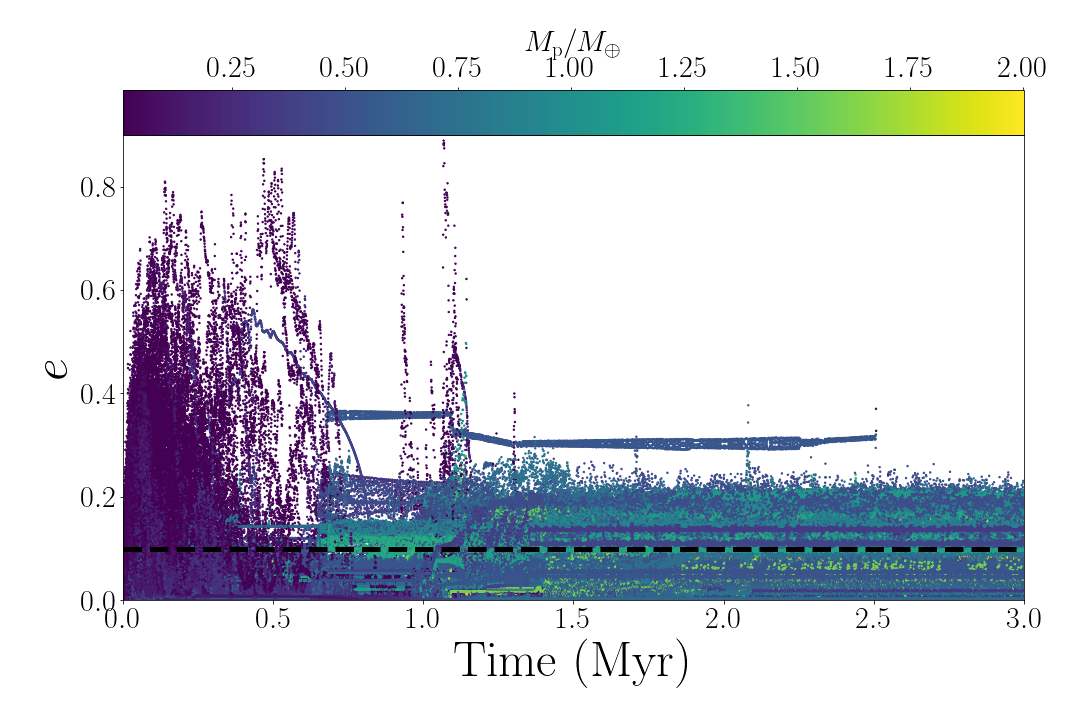}
     \includegraphics[width=\columnwidth]{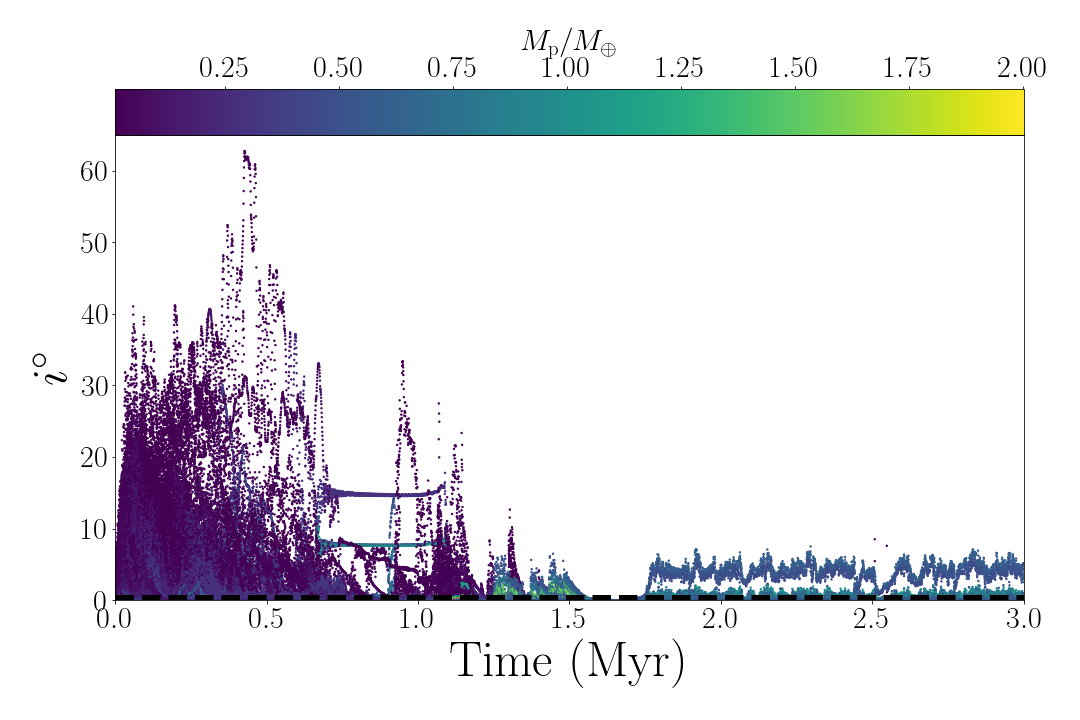}
    \caption{The eccentricity $e$ (top) and inclination $i^{\circ}$ (bottom) of all the bodies in all 24 runs.  The black dashed line marks the initial value of $e$ and $i^{\circ}$ for the starting bodies and the color corresponds to the mass of the body.}
    \label{fig:eccs_and_incs}
\end{figure}

Pebble accretion efficiency increases with the eccentricity of the accreting body until the body moves faster than the pebbles \citep{Liu2018}.  Thus, the eccentricity evolution of the bodies can have significant effects on the final planetary system.  Figure \ref{fig:eccs_and_incs} shows the $e$ and $i$ evolution for all the bodies in the 24 runs.  The black dashed lines mark the initial values for the starting bodies of $e=0.1$ and $i=0.5^{\circ}$.  The color corresponds to the mass of the body.  While our starting eccentricity is larger than what is commonly used in $n$-body models, it is worth noting that the commonly adopted smaller value of $e=0.01$ comes from studies of the solar system where bodies are distributed across the whole range of the starting disc \citep{Chambers1999}.  On the contrary, in our formation channel all bodies start in a narrow annulus just exterior to the ice line and there is a relatively large number density of bodies.  \cite{Schoonenberg_2019} modeled planetesimal formation just exterior to the ice line and found that shortly after formation, the bodies experience body-body scattering which increases the eccentricity distribution of the bodies exterior to the ice line \citep[see][Fig.~4]{Schoonenberg_2019}.  Similar excitation of eccentricity was also found when the planetesimals are formed in narrow axisymmetric dust filaments driven by the streaming instability \citep{Schafer2017}.  Motivated by these findings, we choose to initialize our starting bodies with a larger $e$ than what is commonly used.

From Figure \ref{fig:eccs_and_incs} we can see the bodies in our simulations also experience a high degree of scattering at the start of the simulation that increases both the eccentricity and inclination of the bodies.  At $50 \, \rm Kyr$ the average eccentricity and inclination are $e=0.15$ and $i=6.1^{\circ}$, both larger than our initial values.  Later, the orbits become dampened as their mass grows through pebble accretion and these larger bodies interact with the gas.  At $500 \, \rm Kyr$, the average eccentricity and inclination are $e=0.03$ and $i=0.6^{\circ}$.  Once bodies begin to reach planet size, interactions with the planets can re-excite orbits. The bodies at the end of our simulations have average values of $e=0.07$ and $i=0.04^{\circ}$ and the more massive bodies have less excited orbits than the smaller bodies (see Table \ref{tab:system_data} for the eccentricity and inclination of the final planets).

\subsection{Collision history and formation timescales}

\begin{figure}
	\includegraphics[width=.5\columnwidth]{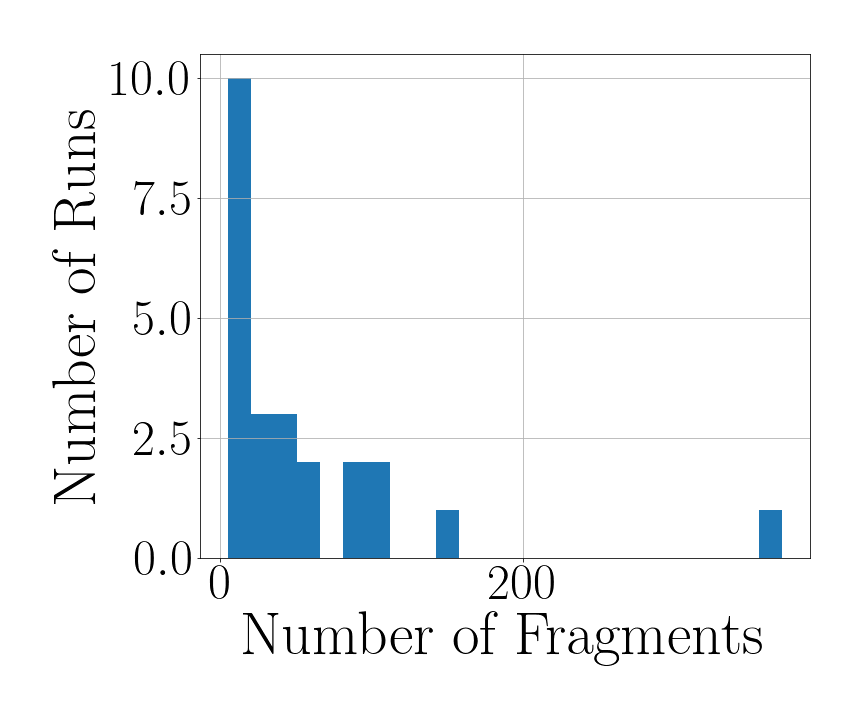}
 	\includegraphics[width=.5\columnwidth]{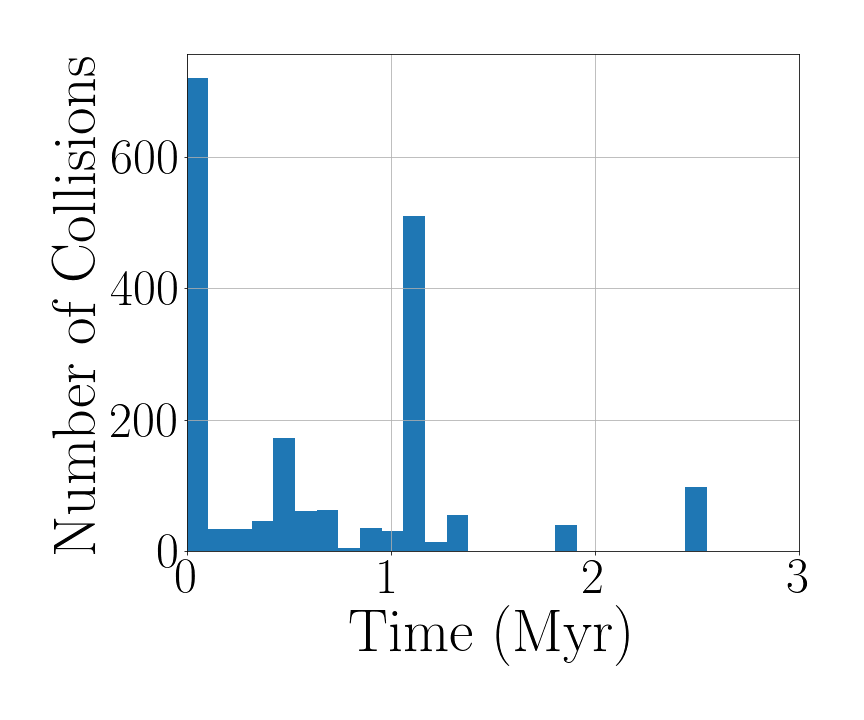}

    \caption{Histogram of the total number of fragments produced in each of the 24 runs we consider (left) and histogram of the time of each collision across each of the 24 runs (right).} 
    \label{fig:frag_hist}
\end{figure}

The collision history is a result of the stochastic behavior of the $n$-body system.  Because we randomise the spatial distribution of the starting planetesimals in each run, we also expect each run to have a different collision history.  We consider the times of the last collision as a proxy for planet formation timescales.  The right panel of Figure \ref{fig:frag_hist} is a histogram of the times for every collision that took place in our 24 runs.  The earliest time of the last collision is $t\sim 1.5 \, \rm Kyr$ in Run38 and Run12 had the last collision at $t \sim 2.5 \, \rm Myr$.  The pileup of collisions at $t \sim 2.5 \, \rm Myr$ are all from Run12.  When considering all 100 runs, Run38 still had the earliest time for the last collision but Run22 had a collision just before $t\sim 3 \, \rm Myr$.  In all 100 runs, pebble accretion continues after the last collision.  In three runs, pebble accretion is still happening just before $3 \, \rm Myr$ and the rest of the runs stop pebble accretion before this time as all the bodies have either reached the PIM or are found interior to a body that has reached the PIM.

Run38 reached its final configuration the earliest and experienced its last collision and ceased pebble accretion before $1 \, \rm Myr$.  While we did not extend our runs in the absence of gas, it is possible that these systems undergo further evolution during and after the gas disk dissipates.  This is expected in systems that do not have stable resonances throughout the system prior to gas disk dissipation. However, the orbital architecture in each of our 24 runs is dominated by strong first order MMRs.

The left panel of Figure \ref{fig:frag_hist} shows a histogram of the total number of fragments produced in each of our 24 systems most similar to T1. Run54 experiences the most fragmentation and produces a total of 371 fragments.  This run experiences a super-catastrophic collision around $1.1 \, \rm Myr$, a relatively late time when the colliding bodies have grown more massive, which results in 339 fragments.  But we view this collision history as atypical and statistically insignificant.  Ten of the runs produce 15 or less fragments.  Run55 and Run100 experience the least fragmentation and produce only five fragments. Relative to the fragmentation that may occur in the solar system, our simulations experience little fragmentation \citep{ChildsSteffen2022}.  However, as discussed below, we find that fragmentation has a significant effect on the multiplicity of the planetary system when fragments are able to grow their mass through pebble accretion.

Fragmentation not only affects the final system architecture and the physical properties of each individual planet, but it can also affect the multiplicity of the system.  In all but one of the 24 runs, at least one planet was seeded by a fragment, that is, a fragment produced from a collision grew its mass by pebble accretion and accreted smaller bodies to become a planet.  On average, three planets in each run are seeded by a fragment.  In Table \ref{tab:system_data} the last three columns show the percentage of the final planet mass that came from pebble accretion, the accretion of fragments, and the accretion of other planetesimals.  We note that in all but three runs the outermost planet is seeded by fragments and grew most of its mass by pebble accretion. 

We further test the effects of fragmentation by performing 100 identical runs where we only resolve collisions with perfect merging.  We find this results in systems with lower multiplicities.  Five out of the 100 runs with perfect merging returned six planets, and the rest of the runs resulted in fewer than six planets.  The mass distribution of the planets with and without fragmentation are similar.  This finding suggests that collisional fragmentation is an important process for systems with high terrestrial planet multiplicities.

\section{Composition and planetary interior structure}\label{sec:Composition}

\begin{figure}
	\includegraphics[width=.99\columnwidth,height=.25\textheight]{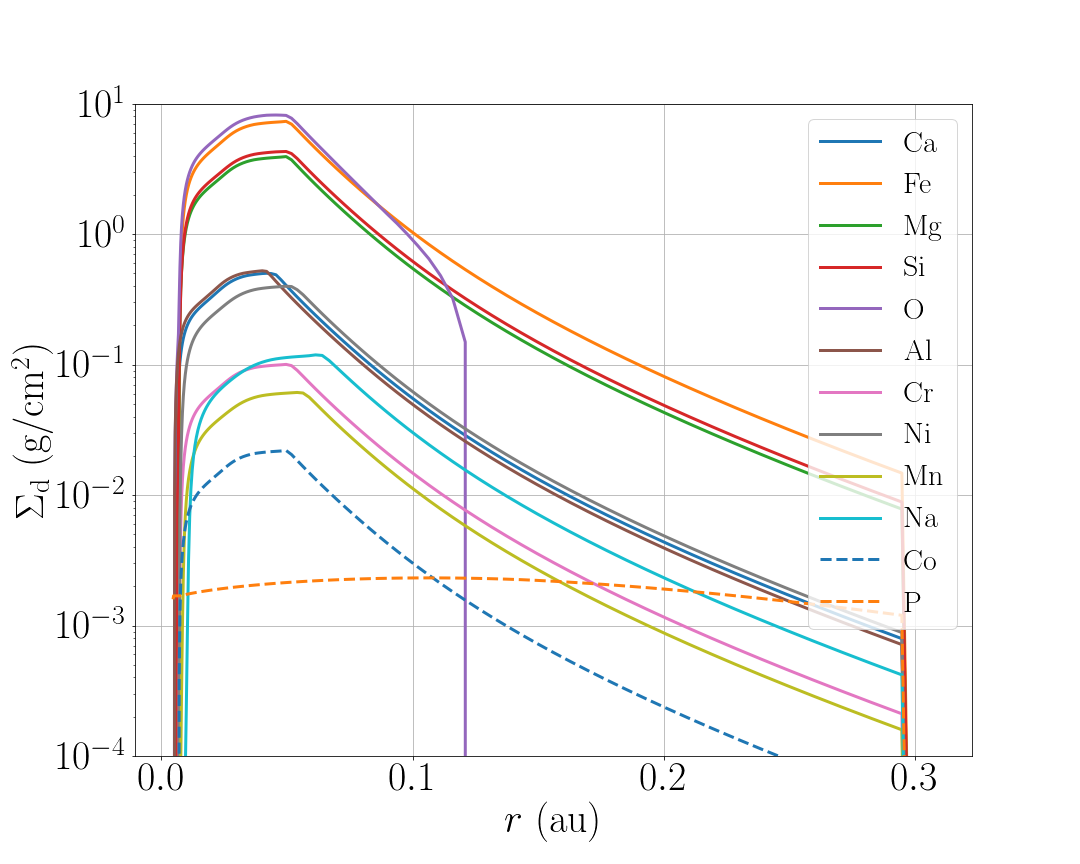}
    \caption{The surface density of 12 elements over 0.01--0.3\,au.  This data is re-scaled from the expected dust profiles around a Sun-like star to be consistent with the size of the disc around a T1-like star.} 
    \label{fig:abundances}
\end{figure}

The T1 planets have observed bulk densities that are consistent with rocky worlds and water mass fractions (WMFs) less than $20\%$ \citep{Agol_2021}.  There are degeneracies between the assumed interior structure and the observed bulk density that result in different WMF predictions for different interior models.  \citet{Acuna2021} model the T1 planets with a coupled atmosphere-interior model and find WMFs for the outer four planets of 9-12\%. Their predicted WMF at a given CMF are 1$\sigma$ higher than reported in \citet{Agol_2021} as they use a less compressible water equation of state. They also find T1-d could have a condensed water layer rather than the water vapor atmosphere assumed in \citet{Agol_2021}.
In this section, we discuss the WMFs along with the elemental compositions of the T1 analogs found in our planet formation models and their implications for the structure of the T1 planets.

Figure \ref{fig:abundances} shows the output of our condensation code for the 12 main elements we focus our analysis on: O, Fe, Si, Mg, Al, Ca, Ni, Na, Cr, Mn, Co, and P.  We choose these elements so that we may make a direct comparison to Earth's compositions as deduced by \cite{MCDONOUGH1995}.  These data are taken from our dust condensation code at five times the characteristic evolution timescale used in \cite{Li2020}, a total of $t\sim 130 \, \rm Kyr$, and used to initialize the composition of the bodies at the start of our $n$-body simulations. The total time of $t\sim 130 \, \rm Kyr$ corresponds to the dust condensation simulations that best match the relative elemental abundances of the solar system's terrestrial planets and the CM, CO, and CV chondrites \citep{Li2020}. Dust condensation is thought to be complete before planetesimal formation so this stage is assumed to take place prior to our $n$-body simulations.  We use these data, along with the assumption that a fraction of the solid material at and exterior to the ice line is water-ice, to set the starting compositions of our bodies and for tracking composition change from pebble accretion.  The dust condensation code does not follow the evolution of the ices and so we must make assumptions about what the intial water ice fraction is for the starting bodies.  

\begin{figure}
            \includegraphics[width=\columnwidth]{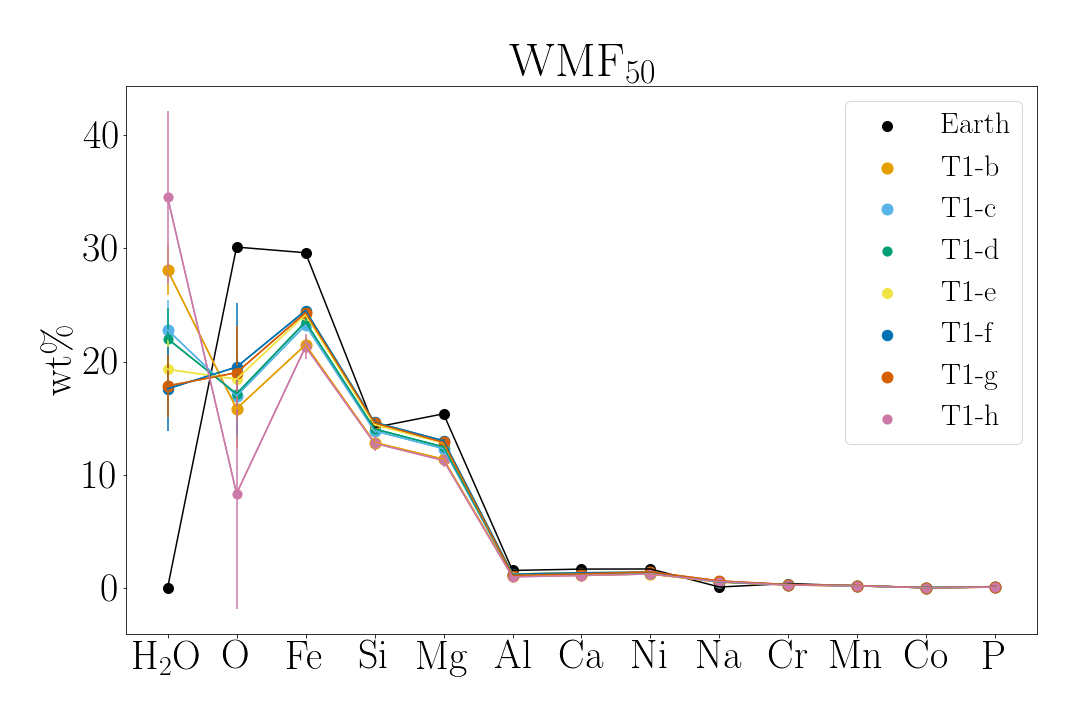}
		\includegraphics[width=\columnwidth]{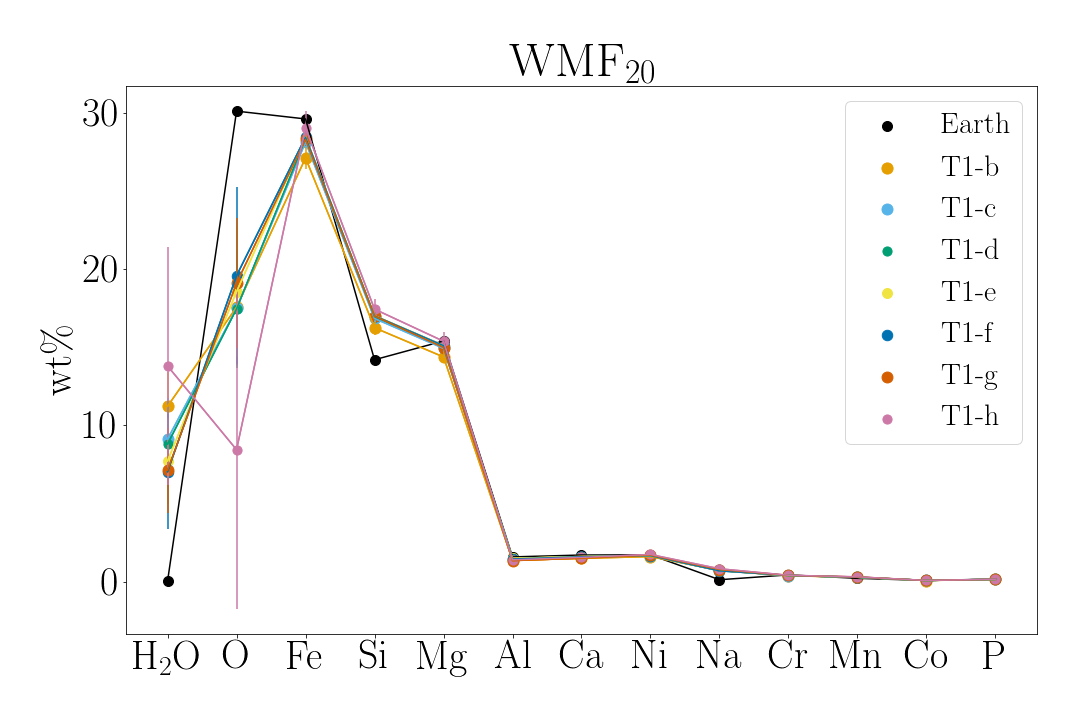}
    \caption{The average $\pm 1 \sigma$ wt\% of water and the 12 elements we consider for each of our average simulated T1 planets.  Earth's values are shown in black. The bodies that begin with a WMF of 50\% are shown in the top panel and the bodies that begin with a WMF of 20\% are shown in the bottom panel.}
    \label{fig:planet_comps}
\end{figure}

Table \ref{tab:system_data} lists the final WMF and elemental abundances for each planet in our 24 runs after running our composition tracking code. We report values for when the bodies were intialised with a 20\% and 50\% WMF.  We find large variability in all of the elements from planet to planet and run to run.  Such variation highlights the sensitivity of the final planet composition on the formation process.  To make a more direct comparison of our simulated planets to the T1 planets, all of the simulated planets are binned into seven semi-major axis bins with the same number of planets in each bin. Figure \ref{fig:planet_comps} shows the average and $\pm 1 \sigma$ for the final wt\% of water and elements for our seven binned T1 analogs with both initial WMFs.  We show Earth's values in black \citep{MCDONOUGH1995}.

As compared to the Earth, it appears that these planets tend to have a reduced fraction of oxygen, iron, and magnesium, but an enhanced WMF. We find the inner and outer planets have higher WMFs in contrast to the trend of increasing WMF with orbital distance found in \citet{Agol_2021}.
Among them, the outermost planet, T1-h, has the largest WMF at its vicinity which is exterior to the ice line.
T1-h has a wide range of WMF and O abundance because the outermost planets can accrete pebbles from either interior or exterior to the ice line where water and condensed oxygen is either depleted or abundant (see Figure ~\ref{fig:abundances}).

Across all runs, we find large variability in the final WMFs of our simulated planets ranging from less than $1\%$ up to $50 \%$ (see Table \ref{tab:system_data}).  If a planet is seeded by a fragment and grows via pebble accretion interior to the ice line, it has a low WMF and hence becomes a dry planet.  On the other hand, if a planet grows via pebble accretion exterior to the ice line, it has a high WMF and is a water world.

All of the average WMFs in our T1 analogs are in excess of that of Earth.  The WMF$_{50}$ values are not consistent with terrestrial worlds, but with water worlds.  Significant volatile loss is thought to take place throughout the planet formation process from impacts, irradiation, and green house effects \citep{Luger2015}.  If we were to model volatile loss, the final WMFs would be lower than what we report here.  However, it is unclear how much devolitisation the T1 planets underwent throughout the formation process.  We examine this more in the following section.

We use our average binned values for elemental compositions of T1 analogs to inform planet interior models with the planet structure code \textsc{Magrathea} \citep{Huang2022}\footnote{\textsc{Magrathea} can be accessed at \url{https://github.com/Huang-CL/Magrathea}}.  \textsc{magrathea} assumes a fully differentiated planet with an iron core, silicate mantle, and water liquid or ice hydrosphere.  After specifying the mass of each layer, the code solves the equations of hydrostatic equilibrium and returns the radius for the planet.  We feed \textsc{magrathea} the seven average binned values for the iron mass fraction for the core, the average binned value of the WMF for the hydrosphere, and place all remaining elements into the mantle.   We use the observed masses of the T1 planets, the default equations of states and phase diagrams in \textsc{magrathea}, and the null-albedo equilibrium temperatures as the start of the adiabatic temperature gradient. We present solutions for a three-layer model (core, mantle, and hydrosphere) here and solutions for a two-layer model (core and mantle) in Section \ref{sec:dry}.  We use a 300 K surface temperature for planets b and c and do not model the water-vapor atmosphere required by their high equilibrium temperatures if they do indeed have a water surface layer (see \citet{Acuna2021} for an atmosphere model of T1-b and T1-c).

Table \ref{tab:WMFs} shows the average binned values for the core mass fraction (CMF), mantle mass fraction (MMF), WMF, and the planet radius as determined by \textsc{magrathea} for a three-layer model for the runs that begin with a 50\% and 20\% WMF. We also include the observed radii ($R_{\rm O}$) from \cite{Agol_2021} for comparison.  The calculated radii that begin with a WMF of 50\% are much larger than observations of the T1 planets. The radii of the T1 analog bodies that begin with a WMF of 20\% agree much better with observations.  As a result, we suggest that either the starting Moon-sized bodies have an initial WMF closer to 20\%, or extreme volatile loss of the planets takes place throughout the planet formation process (see Section \ref{sec:radii} for a discussion of this).

Next, we use the average binned values of iron and silicate from the initial 50\% WMF to find the WMF needed to match the observed masses and radii of the T1 planets. We use 5000 draws of the correlated masses and radii for each planet found in the \citet{Agol_2021} pipeline\footnote{Masses and radii obtained through \url{https://github.com/ericagol/TRAPPIST1_Spitzer}} rather than fitting distributions to the reported median and standard errors. Table \ref{tab:obsWMF} shows the Fe/Si molar ratios (the same abundances are used to find the CMF and MMF in Table \ref{tab:WMFs}) and the WMF results for the outer five planets. We do not include planets b and c in this table as we do not find three-layer solutions that match observations.  The Fe/Si molar ratio is similiar across all seven planets. The value is higher but within 1$\sigma$ of the 0.76$\pm$0.12 used in \citet{Acuna2021} derived from observed stellar abundances in stars similar to T1.

This shows the WMFs need to be reduced down to about 6\%, 4\%, 6\%, 8\%, and 9\% for planets d-h, respectively, to obtain the same radii of the T1 planets. These reductions imply that the planets that form from planetesimals with a 50\% WMF must lose approximately 75\% of their water throughout the formation process in order to match the observed planet densities.  The planet analogs that begin with a 20\% WMF have Fe/Si molar ratios within 0.1\% of the initial 50\% WMF runs but have final WMFs mostly within 2$\sigma$ of the observed inferred WMF in Table \ref{tab:obsWMF}.

\begin{table}
    \centering
    \begin{tabular}{c|c|cccc|c}
  \hline
    {Starting WMF} & {Planet}  & {CMF} & {MMF}  & {WMF} &$R_{\mathrm p}/R_{\oplus}$ &$R_{\mathrm O}/R_{\oplus}$\\
 \hline

\multirow{7}{*}{0.50} & 
   {b} & {0.22} & {0.50} & {0.28}&{1.25} & {1.12}\\
 & {c} & {0.23} & {0.54} & {0.23}&{1.21} & {1.10} \\
 & {d} & {0.24} & {0.54} & {0.22}&{0.84} & {0.79} \\
 & {e} & {0.24} & {0.57} & {0.19}&{0.99} & {0.92}\\
 & {f} & {0.24} & {0.58} & {0.18}&{1.10} & {1.05}\\
 & {g} & {0.24} & {0.58} & {0.18}&{1.18} & {1.13}\\
 & {h} & {0.21} & {0.44} & {0.35}&{0.83} & {0.78}\\

 \hline

\multirow{7}{*}{0.20} & {b} 
& {0.27} & {0.62} & {0.11}&{1.16} & {1.12}\\

 & {c} & {0.28} & {0.63} & {0.09}&{1.13} & {1.10}\\

 & {d} & {0.28} & {0.63} & {0.09}&{0.78} & {0.79}\\
 & {e} & {0.28} & {0.65} & {0.08}&{0.93} & {0.92}\\
 & {f} & {0.28} & {0.65} & {0.07}&{1.04} & {1.05}\\
 & {g} & {0.28} & {0.65} & {0.07}&{1.12} & {1.13}\\
 & {h} & {0.29} & {0.57} & {0.14}&{0.76} & {0.78}\\

    \end{tabular}
    \caption{Core mass fraction (CMF), mantle mass fraction (MMF), water mass fraction (WMF) and resulting radius for the T1 planets using the average bulk compositions from the simulated planets.  We report these properties for starting bodies with a WMF of 0.50 and 0.20.  In the rightmost column are the observed radii ($R_{\rm O}$) for comparison.}
    \label{tab:WMFs}
\end{table}

\begin{table}
    \centering
    \begin{tabular}{c|cccc}
  \hline
    {Planet}  & Fe/Si & CMF & MMF & WMF\\
 \hline\\[-0.8em]
  {d} & {0.84}& {0.283}& {0.656}&{0.061$^{+0.023}_{-0.020}$}\\[0.3em]
  {e} & {0.84}& {0.288}& {0.674}&{0.039$^{+0.021}_{-0.017}$}\\[0.3em]
  {f} & {0.84}& {0.281}& {0.663}&{0.056$^{+0.019}_{-0.015}$}\\[0.3em]
  {g} & {0.84}& {0.274}& {0.651}&{0.075$^{+0.020}_{-0.017}$}\\[0.3em]
  {h} & {0.84}& {0.296}& {0.614}&{0.090$^{+0.047}_{-0.040}$}\\
    \end{tabular}
    \caption{The Fe/Si molar ratio of the T1 analogs from our simulations, the median CMF and MMF implied by the ratio, and the WMF with error needed to match observed mass and radius with error from \citet{Agol_2021}.}
    \label{tab:obsWMF}
\end{table}



\begin{figure}
	\includegraphics[width=\columnwidth]{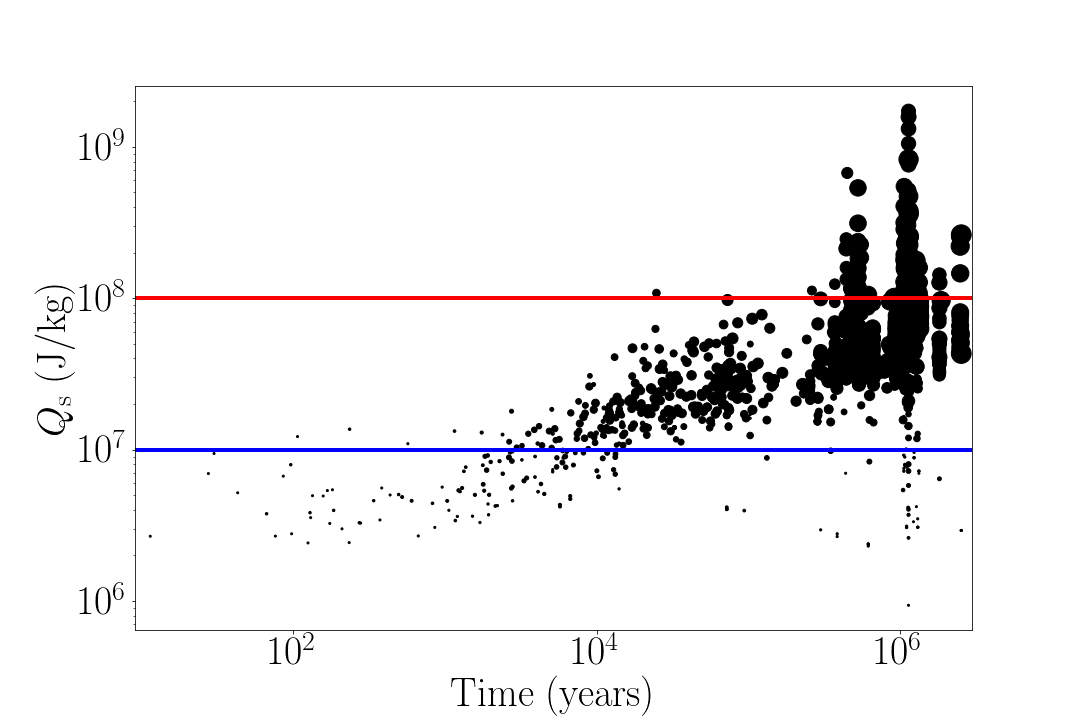}
     \includegraphics[width=\columnwidth]{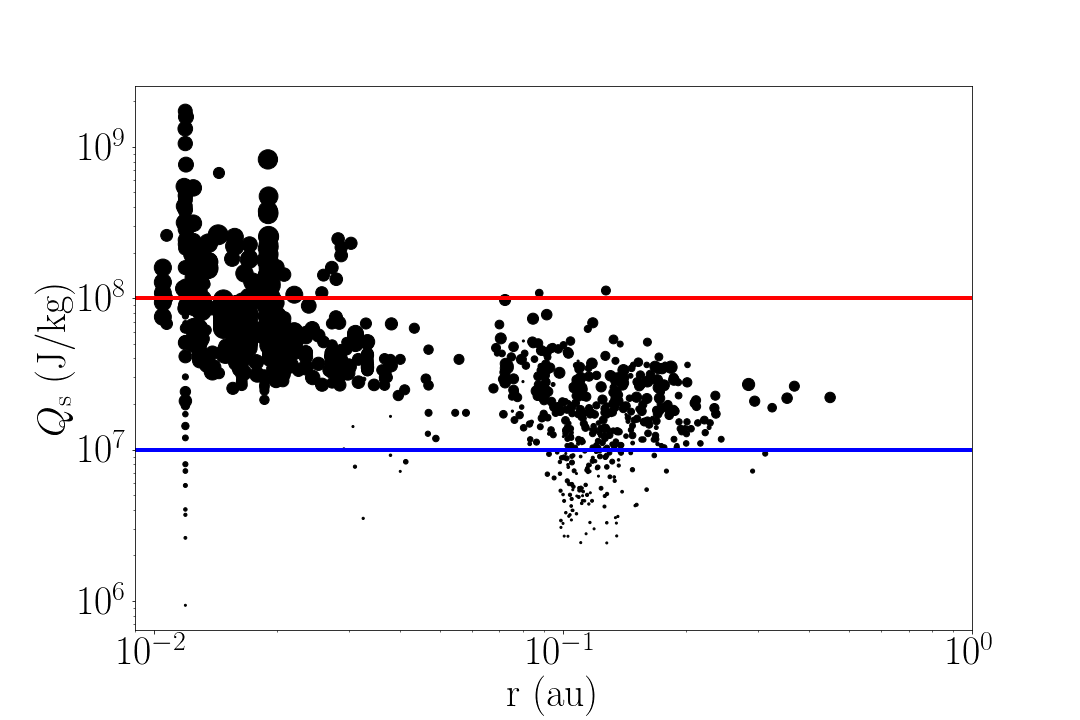}
    \caption{The specific impact energy, $Q_{\rm s}$, of all the collisions the final planets experience as a function of time (top) and as a function of planet orbital radius (botom).  The size of the point is proportional to the total mass of the colliding system.  The blue horizontal line marks the specific energy needed to remove the atmosphere of a planet that has a 1:300 atmosphere-to-ocean ratio and the red line marks the specific energy needed to remove an entire ocean from a similar planet.}
    \label{fig:collision_energies}
\end{figure}

\subsection{Volatile loss}\label{sec:radii}

There are multiple mechanisms that can deplete the planets of their volatile inventory.  Volatile loss can take place prior to the planet formation process in the nebulae as the result of chemical interactions between gas and dust, and the formation of chondrules \citep{Cassen1996, Hirschmann2021}.  Small pebble may lose some of their volatile reserves through pebble abblation \citep{Coleman2019}.  Later, small planetesimals can lose volatiles as they accrete smaller bodies, which heats the body and leads to differentiation \citep{Norris2017, Young2019}.  As the bodies grow into larger embryos which exceed the PIM, planetary growth proceeds via core accretion.  The larger bodies collide with one another in giant impacts, which leads to even further volatile loss \citep{Genda2005, Schlichting2018}.  Lastly, after the final terrestrial planets have formed, the planets may lose their atmospheres through photoevaporation from the host star \citep{Lammer2003}, core-powered mass-loss where heat from the planet's core is thermally transferred to the planet surface and evaporates the atmosphere \citep{Ginzburg2018} or, they may lose volatiles from runaway Greenhouse effects \citep{Luger2015}.

Instead of tracking the multiple ways in which the volatiles may be lost throughout (and after) the formation process, we artificially adjust the WMF while keeping the refractory abundances constant such that the final WMF are similar to those inferred from observations, as described above.  However, we consider the collision history to get a sense of the frequency of giant impacts the T1 planets experienced.

The planets that begin with a 20\% WMF result in planets with radii and densities similar to observations of the T1 plants.  Our simulated planets b, c and e are slightly larger than the observed T1 planets and so these planets would need to undergo slight volatile loss to reproduce the observed radii.  Planets b and c are the largest planets in the system and they do experience collisions throughout the formation process which may be a source of volatile loss.  Because a 20\% WMF reproduces planets with similar radii to the T1 planets without requiring appreciable volatile loss, the starting bodies that formed the T1 planets likely had no less than a 20\% WMF on average.  However, as noted previously T1-b and T1-c may have water vapor atmosphere which would require an even lower water mass fraction \citep{Acuna2021}.

Since our simulations begin after the formation of planetesimals, giant impacts are the main mechanism for removing volatiles at this stage in the planet formation process.  \cite{stewart2014atmospheric} showed that when the specific energy, $Q_{\rm s}$, of a collision between two bodies is more than $10^8 \, \rm J/kg$, it can strip an entire ocean of water on a planet that has an atmosphere-to-ocean ratio of 1:300.  This same planet can have its entire atmosphere striped if it is involved in a collision with $Q_{\rm s} \geq 10^7 \rm J/kg$.  We track the collisions that the simulated planets were involved in to better understand the extent of volatile loss these planets may have experienced. The top panel of Figure \ref{fig:collision_energies} shows $Q_{\rm s}$ versus time for all of the collisions the final planets experienced and the bottom panel shows $Q_{\rm s}$ versus planet orbital radius.  The blue horizontal line marks the specific energy needed to remove the atmosphere of a planet that has a 1:300 atmosphere-to-ocean ratio and the red line marks the specific energy needed to remove an entire ocean from a similar planet.

Two-thirds of our 24 runs had a final planet that experienced at least one ocean evaporating impact.  Of these 16 runs, 14 runs had fewer than 10, Run37 experienced 17, and Run54 experienced 42 ocean evaporating impacts.  Run54 is also the system that experienced the most fragmentation which created a chain reaction of giant impacts.  This can be seen in Figure \ref{fig:collision_energies} near $\sim 1\, \rm Myr$.

All 24 runs experienced atmosphere stripping collisions.  The runs experienced anywhere from 12 up to 83 atmosphere stripping impacts with an average of 36 such collisions.  Run37, which experienced 17 ocean evaporating collisions had the most atmosphere stripping collisions.  In this run, we would expect such excessive volatile loss to result in dry planets. We observe that the ocean evaporating impacts take place in the inner regions of the disc where orbital speeds are higher and more massive planets are found.  As a result, the innermost planets are most susceptible to ocean evaporating impacts which might indicate a lower volatile content for these planets.  However, eight of our simulated runs experience no ocean stripping events and no more than 12 atmosphere stripping events, which suggests the planets may experience little volatile loss from giant impacts.  The atmosphere stripping impacts can be found throughout regions of the disc but are most commonly found around the ice line.  Collisions with smaller impact energies may not result in significant volatile loss and may even be a vehicle for volatile transport \citep{Wakita2019}.  More detailed modeling of volatile loss/gain in collision processes in planet formation scenarios is needed in order to better constrain final volatile budgets.  From Table \ref{tab:system_data}, however, we see that many of the smallest planets in a run form almost all of their mass through pebble accretion.  These planets are not involved in any collisions and so we do not expect the smaller planets to lose any volatiles through collisional processes.

 

\subsection{Desiccated Planets}\label{sec:dry}

While the slight under-density of the T1 planets compared to Earth suggests a volatile layer and our formation models have led to planets with high WMFs, the volatile loss mechanisms discussed above could lead to desiccated planets devoid of volatiles. Recently, JWST observed thermal emission from secondary eclipses of T1-b \citet{greene2023thermal}. The measured temperature of T1-b supports the planet having no atmosphere counter to the water-vapor atmosphere which would result from water-rich surface. We discuss here small core interior solutions to the observed density of the T1 planets which match our composition.

As described above, we find mass fractions of core and mantle with two-layer models in \textsc{Magrathea} which match 5000 draws of the observed masses and radii of the T1 planets from \citet{Agol_2021}. A number of draws of mass and radius do not have solutions assuming only two layers and require a volatile layer. The dessicated CMFs using our simulated Fe/Si ratios (CMF$_{\rm D}$), the CMFs for two-layer planets that match observations (CMF$_{\rm O}$), and the percent of draws with solutions are shown for the seven planets in Table \ref{tab:CMF}.

\begin{table}
    \centering
    \begin{tabular}{c|ccc}
  \hline
    {Planet}  & CMF$_{\rm O}$ & CMF$_{\rm D}$ & Solution \%\\
 \hline\\[-0.8em]
  {b} & 0.20$^{+0.06}_{-0.05}$ & {0.30$^{+0.02}_{-0.02}$} & 92.0\\[0.3em]
  {c} & 0.20$^{+0.05}_{-0.05}$ & {0.30$^{+0.02}_{-0.02}$} & 96.6\\[0.3em]
  {d} & 0.11$^{+0.05}_{-0.05}$ & {0.30$^{+0.02}_{-0.02}$} & 96.2\\[0.3em]
  {e} & 0.18$^{+0.05}_{-0.04}$ & {0.30$^{+0.004}_{-0.01}$} & 99.5\\[0.3em]
  {f} & 0.14$^{+0.04}_{-0.04}$ & {0.30$^{+0.02}_{-0.01}$} & 98.8\\[0.3em]
  {g} & 0.10$^{+0.04}_{-0.04}$ & {0.30$^{+0.01}_{-0.01}$} & 96.5\\[0.3em]
  {h} & 0.05$^{+0.04}_{-0.03}$ & {0.33$^{+0.06}_{-0.04}$} & 45.0\\
    \end{tabular}
    \caption{The inferred CMF from observations (CMF$_{\rm O}$) and the dessicated CMF for our T1 analogs (CMF$_{\rm D}$). CMF$_{\rm O}$ is the CMF needed to match observed density if the planet has only 2-layers and is fully differentiated. CMF$_{\rm D}$ corresponds with the CMF in Table \ref{tab:WMFs} if all the water is removed.  We also list the percent of draws of observed density that can be solved with a two-layer model.}
    \label{tab:CMF}
\end{table}

CMF$_{\rm D}$ is found from the CMFs in Table \ref{tab:WMFs} where CMF$_{\rm D}$=CMF/(1-WMF) and is nearly identical for all planets except for T1-h. For T1-b, T1-c and T1-e, 2$\sigma$ uncertainties of CMF$_{\rm O}$ and the CMF$_{\rm D}$ of our T1 analogs overlap. However for the remaining four planets, the desiccated CMFs for our T1 analogs are 30-33\% which is significantly larger than the CMF inferred from observations for these planets. T1-h needs the smallest core to match observations with a 5\% CMF$_{\rm O}$ . Our dessicated T1-h analog has the largest CMF$_{\rm D}$, 33\%, indicating that all of the iron differentiates into a pure iron core. 

To match our iron weight percentages from formation, the outer T1 planets would need a large amount of the iron in the mantle which we previously assumed to be pure magnesium silicate. However, adding iron to the mantle would increase the density of the mantle \citep{Bejina2019}. If we hold total mass constant, the CMF would need to decrease further with iron in the mantle to match the observed radius. \citet{Agol_2021} found a mass-radius line passes through all seven T1 planets for a core-free composition. For this model they used an abundance ratio for Fe/Si/Mg/O of 29.2/17.3/15.3/38.2 wt\%. The composition of our T1 analogs after the removal of water is on average 30.0/18.1/16.1/21.2 wt\% of Fe/Si/Mg/O. While the Fe/Si/Mg we find coincides well with the core-free composition, a core-free planet requires a high oxygen abundance not seen in our formation models. All of the oxygen weight percents are lower than the 38\% needed to oxidize the iron in the mantle.

Other interior models may also fit our compositions and the observed density of the T1 planets. The CMF can be increased by assuming a liquid core or by putting lighter elements in the core. However, a liquid core only increases the inferred CMF of T1-f from 14 to 15\%. In addition, a promising area of research not investigated here is the incorporation of water into the mantles of the T1 planets. A hydrated mantle could link our formation mechanisms with the observed lack of atmosphere on T1-b. However, \citet{Shah2021} found mantles of 1 M$_\oplus$ planets can store 1-2\% of their mass in water within the mantle which can change the radius by approximately 0.5\% from a dry model. In comparison a 1\% change in CMF changes the radius of the planet by 0.2\%. A change in the assumed CMF is a much larger effect. To match both our formation models and the observed densities, T1-b and T1-c could be desiccated while the outer planets most likely need a significant volatile layer.

\section{Discussion}\label{sec:discussion}

\cite{Vangrootel2018} modeled the stellar evolution of T1 and found that the current luminosity of T1 is $\sim 5 \times 10^{-4} \, L_{\odot}$ and the luminosity of the star at $10 \, \rm Myr$ is $\sim 0.01 \, L_{\odot}$. However, the stellar luminosity of M-dwarfs in their pre-main sequence stages should be orders of magnitudes larger.  \cite{Ramirez2014} modeled the luminosity of a pre-main sequence M8 star and found that when the star was $\sim$1\,Myr old its luminosity was $\sim$0.05\,$L_{\odot}$ from which point it continually dimmed until it reached the main sequence stage and its current luminosity over the course of $\sim$1\,Gyr. 

\begin{figure}
	\includegraphics[width=\columnwidth]{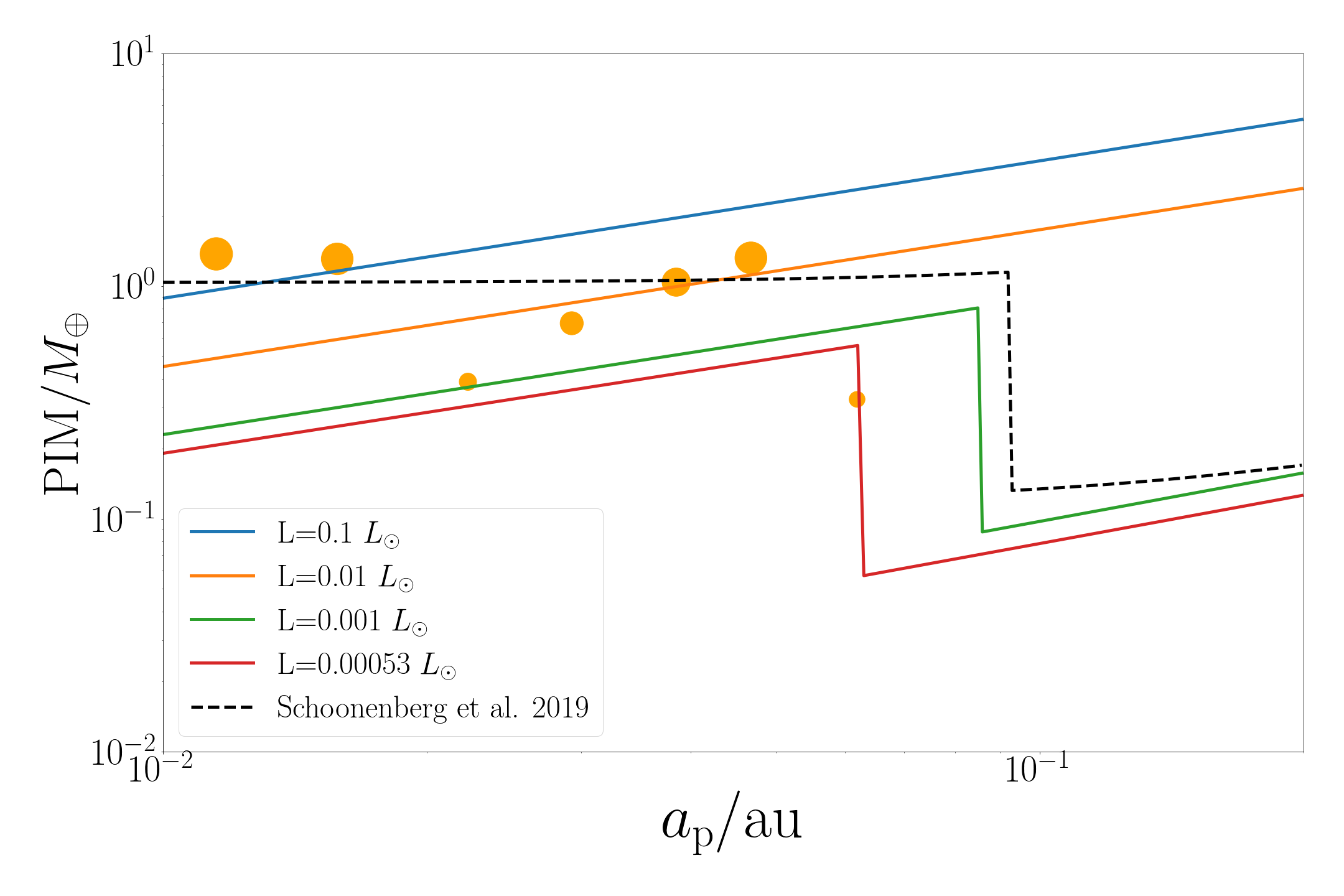}
    \caption{PIM as a function of four different values for stellar luminosity using the  temperature profile for the MMSN which assumes disc heating is dominated by stellar irradiation (solid lines) and also for a temperature profile that accounts for stellar irradiation and viscous heating (black dashed line).  The T1 planets are shown by orange points.}
    \label{fig:PIMvL}
\end{figure} 

The PIM is proportional to the temperature of the disc which in turn, depends on the luminosity of the star and the viscosity of the disc.  Figure \ref{fig:PIMvL} shows the PIM as a function of four different values for luminosity using the \cite{Hayashi1981} temperature profile for the MMSN, Eq. \eqref{eq:Temp_profile}, which assumes disc heating is dominated by stellar irradiation.  We also show the PIM for the temperature profile used in \cite{Schoonenberg_2019} which accounts for stellar irradiation and viscous heating.  We plot the T1 planets in orange.  We see that the ice line and the luminosity change the PIM, which should affect the subsequent evolution of the system.  As the luminosity decreases in time, the PIM also decreases.  The evolution of the stellar luminosity, and thus the PIM, is important to capture when modeling the formation of the T1 planets.  As the PIM decreases with stellar luminosity and time, this may indicate that planets T1-d and T1-h formed at a later time.  Since fragmentation extends the planet formation process and we find that fragments are capable of seeding an entire planet, a fragment produced at a later time which grows by pebble accretion in a relatively depleted gas disc could explain the current masses of T1-d and T1-h.
On the other hand, we note that the other T1 planets appear to follow PIM from an MMSN profile and a specific luminosity.

Nevertheless, accurate temperature profiles are necessary for determining the location of the ice line which also strongly affects the PIM. We adopt a constant, intermediate value for luminosity in our temperature profile and assume a constant location of the ice line which results in planet masses smaller than the observed planet masses for planets e-g.  We assume that the ice line is the location in the disk where the temperature is $170 \, \rm K$. However, \cite{Unterborn2018} found that in the denser protoplanetary discs that exist around M-dwarfs, the ice line is more likely to exist in the region of the disc where the temperature is $212 \, \rm K$.  Future work that implements more accurate temperature profiles in time is necessary to better understand the formation of the T1 system, particularly the role of pebble accretion in the formation process and may more accurately reproduce the observed masses of the T1 planets.  However, a cooling disc alone cannot explain the reversed mass distribution as this would imply planets d-g would need to form from the outside in.  Even though the ice line would move inwards with a cooling disc, it seems unlikely it would move as far in as the current orbits of planets d,e,f or perhaps even g.  An additional explanation for the low mass of T1-h though, would be if the ice line moved interior to T1-h near the time T1-h reached its current mass, thus ceasing pebble accretion and the continued growth of the planet.

The PIM significantly affects the evolution of the system, particularly the mass distribution and thus, migration rates of the planets.  \cite{Bitsch_2018} derived a PIM expression from 3D hydrodynamical simulations while our adopted expression from \cite{Ataiee2018} is derived from 2D hydrodynamical results.  \cite{Ataiee2018} compared their results to \cite{Bitsch_2018} and found overall good agreement however, their PIM values are a factor of 1.5-2 times smaller than those of \cite{Bitsch_2018}.  This may be attributed to the 3D nature of the \cite{Bitsch_2018} simulations where gap carving is more difficult to achieve than in the 2D case.  Exploring how the 3D PIM expression of \cite{Bitsch_2018} affects our results is another area of interest for future work.

Our disc evolution only considers mass loss from accretion onto the central star.  This leads to relatively low rates of mass loss in the disc which may contribute to the inability of our model to reproduce the higher order resonances where the inner planets are found in.  \cite{Ogihara2022} showed that mass loss from photoevaporation and disc winds rapidly deplete the disc mass in the T1 system which results in fast then slow migration of the planets, as well as a late expansion of the orbits due to the clearance of the inner cavity.  The parameters we chose for our disc model result in relatively modest migration rates.  This results in the observed first order MMRs but traps the planets in a three-body resonance not found in the T1 system and does not produce the MMRs of the two inner-most planet pairs. 

A faster disc evolution model that permits planetesimals to quickly reach the PIM and quickly produces short migration timescales may help produce the observed planet orbits.  Fast disc evolution can also help explain the masses of T1-d and T1-h if they were seeded by a fragment and reached their current masses when the disc dissipated, which prevents them from growing to the PIM.  In addition to a disc model that more accurately describes mass loss, a more accurate treatment of the physics in the inner disc cavity may also be needed to produce the observed MMRs between the inner T1 planets \citep{Huang2021}.

While we find that fragmentation is an important mechanism for producing planets as fragments seed planets that grow primarily from pebble accretion, our simulations assume a relatively large minimum fragment mass due to computational limitations.  A smaller minimum fragment mass is likely to affect the planet formation process as more fragments would be produced and each fragment is able to grow its mass through pebble accretion.  We note though that the pebble accretion rate sensitively depends on the accreting mass (Eqs. \ref{eq:mass_growth}-\ref{eq:accretion_radius}).  Additionally, the fragmentation model we use in this study assumes that fragments from a collision are all equally sized and have the same composition.  To better understand the role of fragmentation in the formation process, higher resolution models for fragmentation of differentiated bodies are needed.  In addition to more accurately modeling planet formation, this should help place tighter constraint on the planet composition.  

Volatile loss and gain must be considered in future models in order to reproduce the observed bulk densities of the T1 planets.  Our model, which neglects all volatile loss and any atmospheric accretion, over-estimates water mass fractions when starting with reasonable initial WMFs.  Detailed modeling of various giant impacts that could have produced the Earth's moon indicate that some vaporization of the Earth's mantle took place, for a wide range of impact energies \citep{Carter2020}.  Accurate handling of volatile and mantle loss from giant impacts \citep[e.g.,][]{HM22}, irradiation \citep[e.g.,][]{Lammer2003}, and green house effects \citep[e.g.,][]{Luger2015} should help constrain the composition of the final planetary system.  However, planet encounters with water rich bodies may also result in increasing the planet's water content. Whether an encounter with a water rich body results in net water loss or water gain depends on the specifics of the collision as demonstrated by previous SPH and $n$-body simulations \citep{Burger2020}.  Furthermore, terrestrial planets may directly accrete their atmospheres from the surrounding gas disc which may later be reduced by UV and X-ray radiation from the young host star {\citep{Stokl2015}}.  Detailed modeling with respect to atmospheric gain and loss throughout the formation process is also necessary for placing tighter constraints on the bulk composition and interior structure of the planets.

\section{Conclusions}\label{sec:conclusions}
In this study, we presented a disc evolution and pebble accretion model.  We incorporated this model into \textsc{reboundX} and used our newly developed module to study the formation of the TRAPPIST-1 (T1) planets.  Our model allows for type-I migration and eccentrity and inclination dampening from gas drag in a gas disc.  In our simulations, $0.01 \, M_{\earth}$ bodies began just exterior to the ice line and grew by pebble accretion until the pebble isolation mass (PIM) was reached.  We also modeled collisional accretion and fragmentation of the bodies.  We used results from a dust condensation code to track the composition evolution of the planets.  Using the final compositions of the code and assuming various interior structures, we used the planetary interior structure code \textsc{Magrathea} to obtain radii for our simulated planets.

We reproduced planetary systems that are similar in mass, orbital radius, and multiplicity to the T1 system by numerically modeling planet formation.  We found that Moon-sized bodies quickly grow to the pebble isolation mass exterior to the ice line and migrate inwards at rates that commonly result in first-order MMRs between planetary pairs.  Our model indicates that the largest planets in the inner system likely grew from a combination of embryo, pebble, and fragment accretion and experienced giant impacts, while the smaller planets in the outer system grew mainly by pebble accretion. We also found that fragmentation between larger bodies plays an important role in seeding the smaller planets as the resulting fragments subsequently grow into planets via pebble accretion.

Tracking the formation process of the planets allowed us to place constraints on the initial water content of the bodies at the start of our simulations.  We did not account for any volatile loss but found the inner, larger planets experienced ocean stripping collisions and most planets experienced a few atmosphere stripping collisions.  Assigning the initial bodies a WMF of 50\% resulted in planets with larger radii and lower densities than those observed in the T1 system.  We found that starting bodies with a WMF of 20\% resulted in radii and densities similar to those of the T1 planets.

Using our composition constraints and planet interior structure code we found solutions for a two-layer model for planets b and c.  This, along with the high number of giant impacts the inner planets experienced throughout their formation process, is inline with recent observations that these planets are likely devoid of an atmosphere. However, the two-layer models seem unlikely for most of the remaining outer planets which suggests that these planets have primordial hydrospheres-- an atmosphere and/or a water surface layer.  Our composition constraints also indicated that no planets are consistent with a core-free interior structure.

\section*{Acknowledgements}
We thank Shichun Huang, Rebecca G.\ Martin and Zhaohuan Zhu for useful conversations.  Computer support was provided by UNLV’s National Supercomputing Center. AC acknowledges support from the NSF through grant NSF AST-2107738.
CCY is grateful for the support from NASA via the Astrophysics Theory Program (grant number 80NSSC21K0141), NASA via the Emerging Worlds program (grant number 80NSSC20K0347), and NASA via the Theoretical and Computational Astrophysics Networks program (grant number 80NSSC21K0497).

\section*{Data Availability}\label{sec:Data}
Simulations in this paper made use of the \textsc{rebound} code (Astrophysics Source Code Library identifier {\tt ascl.net/1110.016}) and \textsc{reboundX} (Astrophysics Source Code Library identifier {\tt ascl.net/2011.020}) which can be downloaded freely at \url{http://github.com/hannorein/rebound} and \url{https://github.com/dtamayo/reboundx}, respectively.  The fragmentation code and bulk composition tracking code for \textsc{rebound} (Astrophysics Source Code Library identifier {\tt ascl:2204.010}) may be found at \url{https://github.com/annacrnn/rebound_fragmentation}.  \textsc{magrathea}, the planet interior solver, may be downloaded freely at \url{https://github.com/Huang-CL/Magrathea}.  The data underlying this article will be shared on reasonable request to the corresponding author.



\bibliographystyle{mnras}
\bibliography{main} 


\appendix

\clearpage
\onecolumn

\begin{landscape}
\section{Planet properties}
In Table~\ref{tab:system_data}, we list the properties of the planets in each of our 24 runs (out of 100) that produced a system of at least six planets, which most closely resemble the T1 system.
The second column lists the planet multiplicity (No.) of each system.
The next columns are the properties of each planet, including: mass ($m_{\rm p}$), semi-major axis ($a_{\rm p}$), eccentricity ($e$), inclination ($i$), and relative abundances (by weight) for water (WMF), O, Fe, Si, Mg, Al, Ca, Ni, Na, Cr, Mn, Co, and P.
The slashes (/) separate the results for initial embryos with a starting WMF of either 20\% or 50\%.
The three rightmost columns are the percentages of the planet mass that came from pebble accretion (Peb), framgents (Frag), and embryos (Em), respectively.
\begingroup 
\setlength\tabcolsep{3pt}
\footnotesize

\setlength\LTcapwidth{\textwidth} 

\setlength\LTleft{0pt}            
\setlength\LTright{0pt}
\begin{longtable}{@{\extracolsep{\fill}}|c|c|cccc|ccccccccccccc|ccc|}
\caption{Final properties of the planets in each of our runs that produced six or more planets.}\label{tab:system_data}
\\
\hline
    {Run} & {No.}  & {$M_{\rm p}$} &{$a_{\rm p}$} &{$e$} & {$i$} & {WMF} & {O} & {Fe} & {Si} & {Mg} & {Al} & {Ca} & {Ni} & {Na} &{Cr} & {Mn} &{Co} & {P} & {Peb} & {Frag} & {Em} \\

    & & {$(M_{\oplus})$} & {(au)} & & (deg) & \%$_{20/50}$& \%$_{20/50}$&\%$_{20/50}$ &\%$_{20/50}$&\%$_{20/50}$&\%$_{20/50}$&\%$_{20/50}$&\%$_{20/50}$&\%$_{20/50}$&\%$_{20/50}$&\%$_{20/50}$ &\%$_{20/50}$ &\%$_{20/50}$ &\%&\%&\%\\
\hline
\multirow{9}{*}{Run3} & \multirow{9}{*}{9} 
&{1.5} & {0.012} & {0.03} & {0.0} & {14.0/34.9} & {13.0/11.5} & {27.6/20.3} & {16.5/12.2} & {14.6/10.8} & {1.38/1.03} & {1.51/1.13} & {1.64/1.21} & {0.76/0.55} & {0.39/0.29} & {0.29/0.21} & {0.08/0.06} & {0.15/0.11} & {54.4} & {15.8} & {29.8} \\
&& {1.1} & {0.016} & {0.08} & {0.0} & {9.4/23.6} & {17.5/16.8} & {28.0/22.9} & {16.7/13.7} & {14.8/12.2} & {1.42/1.18} & {1.55/1.28} & {1.66/1.36} & {0.73/0.59} & {0.4/0.33} & {0.29/0.24} & {0.08/0.07} & {0.15/0.13} & {41.0} & {16.8} & {42.3} \\
&& {1.2} & {0.021} & {0.05} & {0.0} & {14.0/35.0} & {9.7/9.3} & {28.9/21.1} & {17.3/12.7} & {15.3/11.2} & {1.41/1.03} & {1.56/1.15} & {1.73/1.26} & {0.8/0.58} & {0.41/0.3} & {0.31/0.22} & {0.08/0.06} & {0.16/0.12} & {35.6} & {17.9} & {46.5} \\
&& {0.5} & {0.034} & {0.04} & {0.0} & {10.5/26.3} & {14.2/14.2} & {28.8/22.9} & {17.3/13.7} & {15.3/12.2} & {1.47/1.19} & {1.61/1.29} & {1.71/1.35} & {0.75/0.58} & {0.41/0.33} & {0.3/0.24} & {0.08/0.07} & {0.16/0.13} & {98.2} & {1.8} & {0.0} \\
&& {0.7} & {0.054} & {0.02} & {0.0} & {9.6/23.9} & {15.7/15.7} & {28.9/23.5} & {17.3/14.1} & {15.3/12.4} & {1.38/1.12} & {1.54/1.25} & {1.73/1.41} & {0.73/0.58} & {0.41/0.34} & {0.3/0.25} & {0.08/0.07} & {0.16/0.13} & {98.4} & {1.6} & {0.0} \\
&& {0.9} & {0.086} & {0.03} & {0.0} & {5.6/13.9} & {19.5/19.5} & {27.9/24.7} & {16.7/14.9} & {14.7/13.1} & {1.34/1.19} & {1.49/1.32} & {1.68/1.49} & {0.81/0.72} & {0.4/0.35} & {0.3/0.27} & {0.08/0.07} & {0.15/0.14} & {98.8} & {1.2} & {0.0} \\
&& {1.0} & {0.097} & {0.03} & {0.0} & {7.5/18.7} & {15.7/15.7} & {28.6/24.4} & {17.2/14.6} & {15.1/12.9} & {1.38/1.18} & {1.53/1.31} & {1.72/1.47} & {0.83/0.71} & {0.41/0.35} & {0.31/0.27} & {0.08/0.07} & {0.16/0.14} & {97.2} & {2.8} & {0.0} \\
&& {0.4} & {0.155} & {0.01} & {0.0} & {20.0/50.0} & {0.1/0.1} & {29.9/18.7} & {18.0/11.2} & {15.9/9.9} & {1.45/0.91} & {1.61/1.01} & {1.8/1.12} & {0.85/0.53} & {0.43/0.27} & {0.32/0.2} & {0.09/0.05} & {0.17/0.1} & {88.5} & {11.5} & {0.0} \\
&& {0.3} & {0.246} & {0.0} & {0.0} & {20.0/50.0} & {0.0/0.0} & {30.0/18.8} & {18.0/11.3} & {15.9/9.9} & {1.45/0.91} & {1.62/1.01} & {1.8/1.12} & {0.85/0.53} & {0.43/0.27} & {0.32/0.2} & {0.09/0.05} & {0.17/0.1} & {95.9} & {4.1} & {0.0} \\
\hline
\multirow{6}{*}{Run12} & \multirow{6}{*}{6} 
& {2.0} & {0.012} & {0.01} & {0.0} & {10.3/25.8} & {16.3/15.6} & {28.0/22.5} & {16.8/13.5} & {14.9/12.0} & {1.43/1.17} & {1.56/1.27} & {1.66/1.33} & {0.73/0.57} & {0.4/0.32} & {0.29/0.23} & {0.08/0.07} & {0.16/0.12} & {8.5} & {21.0} & {70.6} \\
&& {1.3} & {0.016} & {0.02} & {0.0} & {6.3/15.8} & {23.1/22.2} & {27.6/24.4} & {16.5/14.6} & {14.7/13.0} & {1.33/1.18} & {1.49/1.32} & {1.61/1.41} & {0.64/0.55} & {0.39/0.34} & {0.27/0.24} & {0.08/0.07} & {0.15/0.14} & {87.8} & {4.3} & {7.9} \\
&& {0.7} & {0.021} & {0.07} & {0.0} & {5.8/14.6} & {21.2/21.2} & {28.3/25.0} & {17.0/15.0} & {15.0/13.2} & {1.35/1.19} & {1.5/1.33} & {1.71/1.51} & {0.78/0.69} & {0.41/0.36} & {0.31/0.28} & {0.08/0.07} & {0.16/0.14} & {86.6} & {0.0} & {13.4} \\
&& {0.8} & {0.028} & {0.02} & {0.0} & {2.6/6.4} & {22.0/22.0} & {28.0/26.6} & {16.8/15.9} & {14.8/14.1} & {1.35/1.28} & {1.5/1.42} & {1.68/1.6} & {0.82/0.78} & {0.4/0.38} & {0.3/0.29} & {0.08/0.08} & {0.16/0.15} & {98.6} & {1.4} & {0.0} \\
&& {0.4} & {0.044} & {0.02} & {0.0} & {5.0/12.5} & {22.5/22.5} & {28.3/25.5} & {16.9/15.2} & {15.0/13.5} & {1.41/1.28} & {1.56/1.41} & {1.68/1.51} & {0.7/0.62} & {0.4/0.36} & {0.29/0.26} & {0.08/0.07} & {0.16/0.14} & {97.4} & {2.6} & {0.0} \\
&& {0.7} & {0.051} & {0.01} & {0.0} & {8.2/20.6} & {17.4/17.4} & {28.6/24.0} & {17.2/14.4} & {15.2/12.7} & {1.37/1.14} & {1.52/1.27} & {1.71/1.44} & {0.78/0.65} & {0.41/0.34} & {0.31/0.26} & {0.08/0.07} & {0.16/0.13} & {98.3} & {1.7} & {0.0} \\
\hline
\multirow{6}{*}{Run15} & \multirow{6}{*}{6} 
& {0.6} & {0.011} & {0.03} & {0.0} & {13.9/34.6} & {17.5/14.3} & {25.9/19.4} & {15.6/11.6} & {13.7/10.3} & {1.27/0.96} & {1.41/1.06} & {1.55/1.16} & {0.71/0.52} & {0.37/0.28} & {0.28/0.21} & {0.08/0.06} & {0.14/0.11} & {17.5} & {1.8} & {80.7} \\
&& {1.0} & {0.015} & {0.06} & {0.0} & {10.7/26.9} & {15.5/14.8} & {28.1/22.3} & {16.8/13.4} & {14.9/11.8} & {1.39/1.11} & {1.53/1.23} & {1.67/1.33} & {0.76/0.59} & {0.4/0.32} & {0.3/0.23} & {0.08/0.07} & {0.16/0.12} & {27.3} & {1.3} & {71.4} \\
&& {1.1} & {0.02} & {0.05} & {0.0} & {12.2/30.5} & {12.9/12.4} & {28.4/21.8} & {17.0/13.0} & {15.1/11.6} & {1.42/1.1} & {1.56/1.2} & {1.69/1.29} & {0.77/0.58} & {0.4/0.31} & {0.3/0.23} & {0.08/0.06} & {0.16/0.12} & {22.5} & {11.8} & {65.7} \\
&& {0.5} & {0.028} & {0.02} & {0.0} & {8.4/21.1} & {17.1/17.1} & {28.6/23.8} & {17.1/14.3} & {15.2/12.7} & {1.47/1.24} & {1.6/1.34} & {1.69/1.41} & {0.74/0.6} & {0.41/0.34} & {0.3/0.24} & {0.08/0.07} & {0.16/0.13} & {80.1} & {0.0} & {19.9} \\
&& {0.6} & {0.045} & {0.01} & {0.0} & {7.2/18.1} & {19.4/19.4} & {28.5/24.5} & {17.0/14.6} & {15.2/13.0} & {1.48/1.29} & {1.65/1.44} & {1.66/1.42} & {0.68/0.56} & {0.4/0.34} & {0.28/0.24} & {0.08/0.07} & {0.16/0.14} & {98.1} & {1.9} & {0.0} \\
&& {0.7} & {0.059} & {0.01} & {0.0} & {6.0/15.1} & {21.1/21.0} & {28.4/25.0} & {17.0/15.0} & {15.0/13.2} & {1.35/1.18} & {1.5/1.32} & {1.71/1.51} & {0.74/0.64} & {0.41/0.36} & {0.31/0.27} & {0.08/0.07} & {0.16/0.14} & {98.4} & {1.6} & {0.0} \\
\hline
\multirow{9}{*}{Run26} & \multirow{9}{*}{9} 
& {2.0} & {0.012} & {0.12} & {0.0} & {10.8/26.9} & {16.6/15.4} & {27.7/22.1} & {16.6/13.2} & {14.7/11.7} & {1.39/1.12} & {1.52/1.22} & {1.64/1.31} & {0.74/0.58} & {0.39/0.31} & {0.29/0.23} & {0.08/0.06} & {0.15/0.12} & {27.6} & {5.1} & {67.2} \\
&& {1.1} & {0.019} & {0.15} & {0.0} & {5.8/14.5} & {21.7/21.6} & {28.2/25.0} & {16.8/14.9} & {15.0/13.3} & {1.59/1.43} & {1.66/1.49} & {1.64/1.45} & {0.66/0.57} & {0.4/0.35} & {0.28/0.24} & {0.08/0.07} & {0.16/0.14} & {86.6} & {13.4} & {0.0} \\
&& {0.6} & {0.025} & {0.07} & {0.0} & {6.8/17.0} & {20.2/20.1} & {28.4/24.6} & {17.0/14.7} & {15.1/13.1} & {1.53/1.34} & {1.67/1.47} & {1.65/1.42} & {0.67/0.56} & {0.4/0.35} & {0.28/0.24} & {0.08/0.07} & {0.16/0.14} & {83.7} & {0.0} & {16.3} \\
&& {0.7} & {0.033} & {0.05} & {0.0} & {7.2/18.1} & {19.3/19.3} & {28.5/24.5} & {17.1/14.7} & {15.1/13.0} & {1.35/1.16} & {1.51/1.3} & {1.72/1.48} & {0.72/0.61} & {0.41/0.35} & {0.31/0.26} & {0.08/0.07} & {0.16/0.14} & {82.6} & {0.0} & {17.4} \\
&& {0.3} & {0.044} & {0.05} & {0.0} & {0.0/0.0} & {30.6/30.6} & {27.7/27.7} & {16.5/16.5} & {14.8/14.8} & {1.54/1.54} & {1.67/1.67} & {1.59/1.59} & {0.58/0.58} & {0.39/0.39} & {0.26/0.26} & {0.08/0.08} & {0.15/0.15} & {96.8} & {3.2} & {0.0} \\
&& {0.7} & {0.057} & {0.04} & {0.0} & {7.4/18.5} & {18.4/18.4} & {28.4/24.3} & {17.1/14.6} & {15.1/12.8} & {1.36/1.16} & {1.51/1.29} & {1.71/1.46} & {0.76/0.64} & {0.41/0.35} & {0.31/0.26} & {0.08/0.07} & {0.16/0.13} & {98.6} & {1.4} & {0.0} \\
&& {0.8} & {0.075} & {0.04} & {0.0} & {1.1/2.8} & {26.8/26.8} & {27.2/26.5} & {16.3/15.9} & {14.4/14.1} & {1.31/1.28} & {1.45/1.42} & {1.64/1.6} & {0.8/0.78} & {0.39/0.38} & {0.3/0.29} & {0.08/0.08} & {0.15/0.15} & {98.8} & {1.2} & {0.0} \\
&& {0.9} & {0.091} & {0.03} & {0.0} & {0.2/0.6} & {26.4/26.4} & {27.3/27.1} & {16.4/16.3} & {14.4/14.4} & {1.31/1.31} & {1.46/1.45} & {1.64/1.63} & {0.79/0.79} & {0.39/0.39} & {0.3/0.3} & {0.08/0.08} & {0.15/0.15} & {98.9} & {1.1} & {0.0} \\
&& {0.3} & {0.103} & {0.03} & {0.0} & {19.6/49.0} & {1.1/0.9} & {29.7/18.8} & {17.8/11.3} & {15.7/10.0} & {1.44/0.91} & {1.6/1.01} & {1.78/1.13} & {0.84/0.53} & {0.42/0.27} & {0.32/0.2} & {0.09/0.05} & {0.16/0.1} & {95.3} & {4.7} & {0.0} \\
\hline
\multirow{8}{*}{Run29} & \multirow{8}{*}{8} 
& {0.6} & {0.012} & {0.18} & {0.0} & {11.7/29.2} & {19.6/16.9} & {26.1/20.6} & {15.7/12.4} & {13.9/11.0} & {1.31/1.04} & {1.44/1.14} & {1.55/1.22} & {0.7/0.54} & {0.37/0.29} & {0.27/0.22} & {0.08/0.06} & {0.14/0.11} & {21.0} & {16.1} & {62.8} \\
&& {1.5} & {0.016} & {0.15} & {0.0} & {11.2/28.0} & {14.5/13.9} & {28.2/22.2} & {16.9/13.3} & {15.0/11.8} & {1.4/1.11} & {1.55/1.22} & {1.68/1.32} & {0.76/0.59} & {0.4/0.32} & {0.3/0.23} & {0.08/0.07} & {0.16/0.12} & {37.0} & {9.1} & {53.9} \\
&& {0.5} & {0.021} & {0.18} & {0.0} & {8.7/21.6} & {17.2/17.2} & {28.6/23.8} & {17.1/14.2} & {15.2/12.6} & {1.52/1.28} & {1.63/1.37} & {1.68/1.39} & {0.71/0.57} & {0.41/0.34} & {0.29/0.24} & {0.08/0.07} & {0.16/0.13} & {97.9} & {2.1} & {0.0} \\
&& {0.5} & {0.026} & {0.06} & {0.0} & {9.5/23.9} & {15.7/15.6} & {28.5/23.3} & {17.1/13.9} & {15.1/12.3} & {1.45/1.2} & {1.59/1.3} & {1.69/1.37} & {0.75/0.59} & {0.41/0.33} & {0.3/0.24} & {0.08/0.07} & {0.16/0.13} & {78.5} & {2.5} & {18.9} \\
&& {0.6} & {0.034} & {0.09} & {0.0} & {10.5/26.2} & {16.0/15.4} & {28.1/22.5} & {16.8/13.5} & {14.9/11.9} & {1.45/1.18} & {1.57/1.27} & {1.66/1.32} & {0.73/0.57} & {0.4/0.32} & {0.29/0.23} & {0.08/0.07} & {0.16/0.12} & {50.8} & {4.6} & {44.6} \\
&& {0.7} & {0.054} & {0.01} & {0.0} & {14.9/37.4} & {7.4/7.4} & {29.3/20.9} & {17.6/12.5} & {15.5/11.1} & {1.41/1.01} & {1.57/1.12} & {1.76/1.26} & {0.84/0.6} & {0.42/0.3} & {0.32/0.23} & {0.09/0.06} & {0.16/0.12} & {61.4} & {38.6} & {0.0} \\
&& {0.9} & {0.086} & {0.01} & {0.0} & {5.4/13.4} & {19.7/19.7} & {27.8/24.8} & {16.7/14.9} & {14.7/13.1} & {1.34/1.19} & {1.49/1.33} & {1.67/1.49} & {0.81/0.73} & {0.4/0.36} & {0.3/0.27} & {0.08/0.07} & {0.15/0.14} & {98.8} & {1.2} & {0.0} \\
&& {0.4} & {0.137} & {0.01} & {0.0} & {19.9/49.8} & {0.1/0.1} & {30.0/18.8} & {18.0/11.3} & {15.9/9.9} & {1.45/0.91} & {1.61/1.01} & {1.8/1.13} & {0.85/0.53} & {0.43/0.27} & {0.32/0.2} & {0.09/0.05} & {0.17/0.1} & {96.9} & {3.1} & {0.0} \\
\hline
\multirow{6}{*}{Run37} & \multirow{6}{*}{6} 
& {1.5} & {0.01} & {0.12} & {0.0} & {11.5/28.8} & {16.0/14.7} & {27.6/21.6} & {16.5/13.0} & {14.6/11.5} & {1.4/1.11} & {1.53/1.21} & {1.64/1.28} & {0.73/0.56} & {0.39/0.31} & {0.29/0.22} & {0.08/0.06} & {0.15/0.12} & {28.1} & {44.3} & {27.6} \\
&& {1.9} & {0.015} & {0.17} & {0.0} & {12.2/30.5} & {13.1/12.6} & {28.5/21.9} & {17.1/13.1} & {15.1/11.6} & {1.56/1.24} & {1.65/1.29} & {1.68/1.28} & {0.73/0.54} & {0.4/0.31} & {0.29/0.22} & {0.08/0.06} & {0.16/0.12} & {8.7} & {28.5} & {62.8} \\
&& {0.5} & {0.023} & {0.08} & {0.0} & {6.6/16.6} & {20.0/20.0} & {28.4/24.7} & {17.0/14.7} & {15.1/13.1} & {1.52/1.34} & {1.63/1.43} & {1.66/1.44} & {0.69/0.59} & {0.4/0.35} & {0.29/0.25} & {0.08/0.07} & {0.16/0.14} & {97.7} & {2.3} & {0.0} \\
&& {0.5} & {0.03} & {0.05} & {0.0} & {8.2/20.4} & {17.8/17.8} & {28.6/24.0} & {17.1/14.4} & {15.2/12.8} & {1.52/1.3} & {1.63/1.39} & {1.68/1.4} & {0.7/0.57} & {0.4/0.34} & {0.29/0.24} & {0.08/0.07} & {0.16/0.13} & {97.9} & {2.1} & {0.0} \\
&& {0.6} & {0.04} & {0.04} & {0.0} & {9.6/24.1} & {15.6/15.6} & {28.8/23.4} & {17.2/14.0} & {15.3/12.4} & {1.46/1.2} & {1.6/1.31} & {1.7/1.38} & {0.73/0.58} & {0.41/0.33} & {0.3/0.24} & {0.08/0.07} & {0.16/0.13} & {98.3} & {1.7} & {0.0} \\
&& {0.8} & {0.063} & {0.0} & {0.0} & {6.8/16.9} & {19.7/19.7} & {28.4/24.6} & {17.0/14.8} & {15.0/13.0} & {1.35/1.17} & {1.51/1.31} & {1.71/1.48} & {0.82/0.71} & {0.41/0.35} & {0.31/0.27} & {0.08/0.07} & {0.16/0.14} & {98.5} & {1.5} & {0.0} \\
\hline
\multirow{6}{*}{Run38} & \multirow{6}{*}{6} 
& {0.5} & {0.013} & {0.19} & {0.0} & {7.8/19.5} & {23.4/21.5} & {26.6/22.9} & {15.9/13.7} & {14.1/12.2} & {1.4/1.22} & {1.51/1.31} & {1.56/1.34} & {0.67/0.56} & {0.38/0.32} & {0.27/0.23} & {0.08/0.07} & {0.15/0.13} & {40.8} & {12.3} & {46.9} \\
&& {0.5} & {0.016} & {0.14} & {0.0} & {9.9/24.6} & {15.9/15.6} & {28.3/22.9} & {16.9/13.7} & {15.0/12.2} & {1.43/1.17} & {1.56/1.27} & {1.68/1.36} & {0.75/0.6} & {0.4/0.33} & {0.3/0.24} & {0.08/0.07} & {0.16/0.13} & {58.1} & {10.4} & {31.5} \\
&& {0.6} & {0.02} & {0.15} & {0.0} & {13.6/34.0} & {14.1/12.3} & {27.3/20.3} & {16.4/12.2} & {14.4/10.8} & {1.33/0.99} & {1.47/1.1} & {1.63/1.21} & {0.76/0.56} & {0.39/0.29} & {0.29/0.22} & {0.08/0.06} & {0.15/0.11} & {42.1} & {7.6} & {50.3} \\
&& {0.5} & {0.032} & {0.05} & {0.0} & {7.2/18.1} & {19.4/19.3} & {28.3/24.3} & {16.9/14.5} & {15.0/12.9} & {1.52/1.32} & {1.62/1.41} & {1.66/1.42} & {0.7/0.58} & {0.4/0.34} & {0.29/0.24} & {0.08/0.07} & {0.16/0.13} & {80.3} & {0.0} & {19.7} \\
&& {0.4} & {0.05} & {0.07} & {0.0} & {5.8/14.5} & {21.5/21.5} & {28.5/25.2} & {17.0/15.0} & {15.2/13.4} & {1.38/1.22} & {1.54/1.37} & {1.65/1.46} & {0.66/0.57} & {0.4/0.35} & {0.28/0.25} & {0.08/0.07} & {0.16/0.14} & {97.0} & {3.0} & {0.0} \\
&& {0.7} & {0.058} & {0.03} & {0.0} & {10.2/25.4} & {15.2/15.0} & {28.5/22.9} & {17.1/13.8} & {15.1/12.1} & {1.36/1.09} & {1.52/1.22} & {1.72/1.38} & {0.78/0.62} & {0.41/0.33} & {0.31/0.25} & {0.08/0.07} & {0.16/0.13} & {62.1} & {0.0} & {37.9} \\
\hline
\multirow{11}{*}{Run39} & \multirow{11}{*}{11} 
& {1.3} & {0.011} & {0.05} & {0.0} & {14.1/35.3} & {12.1/10.8} & {27.8/20.4} & {16.7/12.2} & {14.7/10.8} & {1.35/1.0} & {1.5/1.1} & {1.66/1.22} & {0.78/0.56} & {0.4/0.29} & {0.3/0.22} & {0.08/0.06} & {0.15/0.11} & {32.7} & {25.4} & {42.0} \\
&& {1.0} & {0.016} & {0.14} & {0.0} & {7.1/17.6} & {19.8/19.6} & {28.2/24.4} & {16.9/14.6} & {15.0/12.9} & {1.5/1.32} & {1.61/1.41} & {1.66/1.42} & {0.7/0.59} & {0.4/0.34} & {0.29/0.24} & {0.08/0.07} & {0.16/0.13} & {41.3} & {3.2} & {55.6} \\
&& {0.9} & {0.021} & {0.09} & {0.0} & {7.9/19.8} & {18.5/18.4} & {28.4/24.0} & {17.0/14.4} & {15.1/12.8} & {1.53/1.32} & {1.63/1.4} & {1.66/1.4} & {0.7/0.57} & {0.4/0.34} & {0.29/0.24} & {0.08/0.07} & {0.16/0.13} & {60.0} & {28.4} & {11.6} \\
&& {0.6} & {0.027} & {0.04} & {0.0} & {8.9/22.2} & {17.4/17.1} & {28.3/23.4} & {16.9/14.0} & {15.0/12.4} & {1.43/1.2} & {1.57/1.31} & {1.68/1.39} & {0.73/0.59} & {0.4/0.33} & {0.29/0.24} & {0.08/0.07} & {0.16/0.13} & {67.1} & {15.3} & {17.6} \\
&& {0.4} & {0.033} & {0.05} & {0.0} & {0.2/0.6} & {31.5/31.5} & {28.0/27.8} & {16.5/16.4} & {15.0/14.9} & {2.08/2.08} & {1.98/1.98} & {1.51/1.51} & {0.4/0.4} & {0.38/0.38} & {0.23/0.22} & {0.08/0.08} & {0.15/0.15} & {97.4} & {2.6} & {0.0} \\
&& {0.3} & {0.04} & {0.03} & {0.0} & {0.2/0.6} & {31.5/31.5} & {27.9/27.8} & {16.5/16.4} & {15.0/14.9} & {2.01/2.0} & {1.95/1.94} & {1.52/1.51} & {0.42/0.42} & {0.38/0.38} & {0.23/0.23} & {0.08/0.08} & {0.15/0.15} & {96.5} & {3.5} & {0.0} \\
&& {0.4} & {0.046} & {0.03} & {0.0} & {0.5/1.2} & {30.6/30.5} & {28.0/27.7} & {16.6/16.5} & {14.9/14.8} & {1.55/1.54} & {1.73/1.72} & {1.57/1.56} & {0.52/0.51} & {0.39/0.39} & {0.25/0.25} & {0.08/0.08} & {0.15/0.15} & {90.7} & {9.3} & {0.0} \\
&& {0.5} & {0.054} & {0.03} & {0.0} & {3.1/7.7} & {26.6/26.3} & {27.8/26.2} & {16.7/15.8} & {14.7/13.9} & {1.31/1.23} & {1.47/1.38} & {1.66/1.57} & {0.65/0.6} & {0.4/0.38} & {0.29/0.27} & {0.08/0.08} & {0.15/0.15} & {98.0} & {2.0} & {0.0} \\
&& {0.8} & {0.063} & {0.03} & {0.0} & {6.6/16.6} & {20.1/20.0} & {28.3/24.6} & {17.0/14.8} & {15.0/13.0} & {1.35/1.17} & {1.5/1.31} & {1.7/1.48} & {0.81/0.7} & {0.4/0.35} & {0.31/0.27} & {0.08/0.07} & {0.16/0.14} & {83.6} & {3.2} & {13.2} \\
&& {0.9} & {0.082} & {0.02} & {0.0} & {2.4/6.0} & {24.2/24.2} & {27.3/25.9} & {16.4/15.6} & {14.4/13.7} & {1.31/1.24} & {1.46/1.38} & {1.64/1.56} & {0.8/0.76} & {0.39/0.37} & {0.3/0.28} & {0.08/0.08} & {0.15/0.14} & {98.9} & {1.1} & {0.0} \\
&& {1.0} & {0.1} & {0.01} & {0.0} & {2.5/6.4} & {24.1/23.3} & {27.2/26.1} & {16.3/15.7} & {14.4/13.8} & {1.31/1.25} & {1.45/1.39} & {1.64/1.57} & {0.79/0.76} & {0.39/0.37} & {0.3/0.28} & {0.08/0.08} & {0.15/0.14} & {98.9} & {1.1} & {0.0} \\
\hline
\multirow{8}{*}{Run41} & \multirow{8}{*}{8} 
& {1.4} & {0.012} & {0.17} & {0.0} & {9.2/23.0} & {19.7/18.3} & {27.3/22.6} & {16.3/13.5} & {14.5/12.0} & {1.4/1.18} & {1.53/1.28} & {1.61/1.33} & {0.7/0.57} & {0.39/0.32} & {0.28/0.23} & {0.08/0.07} & {0.15/0.13} & {14.8} & {4.1} & {81.1} \\
&& {1.5} & {0.019} & {0.1} & {0.0} & {8.3/20.9} & {19.2/18.5} & {27.8/23.4} & {16.6/14.0} & {14.8/12.4} & {1.45/1.23} & {1.57/1.33} & {1.64/1.38} & {0.71/0.58} & {0.39/0.33} & {0.29/0.24} & {0.08/0.07} & {0.15/0.13} & {50.0} & {12.0} & {37.9} \\
&& {0.5} & {0.025} & {0.1} & {0.0} & {9.5/23.8} & {16.7/16.2} & {28.2/23.1} & {16.9/13.8} & {15.0/12.3} & {1.45/1.2} & {1.58/1.3} & {1.67/1.36} & {0.73/0.59} & {0.4/0.33} & {0.29/0.24} & {0.08/0.07} & {0.16/0.13} & {77.3} & {0.0} & {22.7} \\
&& {0.5} & {0.033} & {0.07} & {0.0} & {8.0/20.0} & {18.0/18.0} & {28.6/24.1} & {17.1/14.4} & {15.2/12.8} & {1.45/1.23} & {1.58/1.34} & {1.69/1.42} & {0.71/0.59} & {0.41/0.34} & {0.29/0.25} & {0.08/0.07} & {0.16/0.13} & {98.0} & {2.0} & {0.0} \\
&& {0.6} & {0.044} & {0.04} & {0.0} & {7.9/19.8} & {18.4/18.3} & {28.6/24.1} & {17.1/14.4} & {15.2/12.8} & {1.51/1.3} & {1.65/1.41} & {1.67/1.4} & {0.69/0.56} & {0.4/0.34} & {0.29/0.24} & {0.08/0.07} & {0.16/0.13} & {98.3} & {1.7} & {0.0} \\
&& {0.8} & {0.069} & {0.03} & {0.0} & {5.0/12.6} & {21.9/21.9} & {28.0/25.2} & {16.8/15.1} & {14.8/13.3} & {1.34/1.2} & {1.49/1.34} & {1.69/1.52} & {0.82/0.74} & {0.4/0.36} & {0.31/0.28} & {0.08/0.07} & {0.16/0.14} & {98.6} & {1.4} & {0.0} \\
&& {0.9} & {0.084} & {0.03} & {0.0} & {3.4/8.4} & {22.7/22.7} & {27.5/25.6} & {16.5/15.4} & {14.5/13.5} & {1.32/1.23} & {1.47/1.36} & {1.65/1.54} & {0.81/0.75} & {0.39/0.37} & {0.3/0.28} & {0.08/0.07} & {0.15/0.14} & {98.8} & {1.2} & {0.0} \\
&& {0.3} & {0.102} & {0.02} & {0.0} & {20.0/49.9} & {0.2/0.1} & {29.9/18.7} & {17.9/11.2} & {15.8/9.9} & {1.45/0.91} & {1.61/1.01} & {1.79/1.12} & {0.85/0.53} & {0.43/0.27} & {0.32/0.2} & {0.09/0.05} & {0.17/0.1} & {96.5} & {3.5} & {0.0} \\
\hline
\multirow{6}{*}{Run46} & \multirow{6}{*}{6} 
& {1.3} & {0.013} & {0.08} & {0.0} & {11.6/28.9} & {16.2/14.7} & {27.4/21.5} & {16.4/12.9} & {14.5/11.4} & {1.35/1.07} & {1.49/1.18} & {1.63/1.28} & {0.75/0.57} & {0.39/0.31} & {0.29/0.23} & {0.08/0.06} & {0.15/0.12} & {21.1} & {3.8} & {75.1} \\
&& {0.8} & {0.016} & {0.11} & {0.0} & {14.9/37.3} & {9.6/8.7} & {28.3/20.3} & {17.0/12.2} & {15.0/10.8} & {1.36/0.98} & {1.52/1.09} & {1.7/1.22} & {0.81/0.58} & {0.41/0.29} & {0.31/0.22} & {0.08/0.06} & {0.16/0.11} & {54.7} & {14.9} & {30.4} \\
&& {0.5} & {0.021} & {0.07} & {0.0} & {7.5/18.7} & {19.2/19.0} & {28.3/24.2} & {16.9/14.4} & {15.0/12.8} & {1.5/1.3} & {1.61/1.39} & {1.66/1.41} & {0.7/0.59} & {0.4/0.34} & {0.29/0.24} & {0.08/0.07} & {0.16/0.13} & {79.0} & {0.0} & {21.0} \\
&& {0.6} & {0.026} & {0.05} & {0.0} & {10.9/27.4} & {13.4/13.4} & {28.9/22.7} & {17.3/13.6} & {15.3/12.1} & {1.45/1.15} & {1.59/1.26} & {1.71/1.34} & {0.76/0.59} & {0.41/0.32} & {0.3/0.24} & {0.08/0.07} & {0.16/0.13} & {79.9} & {20.1} & {0.0} \\
&& {0.6} & {0.041} & {0.02} & {0.0} & {5.9/14.9} & {21.4/21.4} & {28.4/25.0} & {16.9/14.9} & {15.1/13.3} & {1.59/1.43} & {1.68/1.5} & {1.64/1.44} & {0.65/0.56} & {0.4/0.35} & {0.28/0.24} & {0.08/0.07} & {0.16/0.14} & {98.1} & {1.9} & {0.0} \\
&& {0.6} & {0.047} & {0.02} & {0.0} & {8.0/20.1} & {18.1/18.1} & {28.7/24.2} & {17.1/14.4} & {15.2/12.9} & {1.43/1.21} & {1.6/1.36} & {1.67/1.4} & {0.69/0.56} & {0.41/0.34} & {0.29/0.24} & {0.08/0.07} & {0.16/0.13} & {98.2} & {1.8} & {0.0} \\
\hline
\multirow{8}{*}{Run47} & \multirow{8}{*}{8} 
& {1.8} & {0.012} & {0.09} & {0.0} & {12.5/31.2} & {13.4/12.5} & {28.1/21.4} & {16.8/12.8} & {14.9/11.4} & {1.38/1.06} & {1.53/1.17} & {1.67/1.28} & {0.77/0.58} & {0.4/0.31} & {0.3/0.23} & {0.08/0.06} & {0.16/0.12} & {18.6} & {5.8} & {75.5} \\
&& {1.3} & {0.017} & {0.21} & {0.0} & {8.1/20.3} & {19.2/18.7} & {28.0/23.7} & {16.8/14.2} & {14.9/12.6} & {1.36/1.15} & {1.52/1.29} & {1.65/1.39} & {0.7/0.58} & {0.4/0.34} & {0.29/0.24} & {0.08/0.07} & {0.16/0.13} & {30.9} & {2.1} & {67.0} \\
&& {0.6} & {0.022} & {0.1} & {0.0} & {8.2/20.5} & {17.8/17.8} & {28.7/24.1} & {17.2/14.5} & {15.2/12.8} & {1.37/1.15} & {1.53/1.28} & {1.71/1.44} & {0.71/0.58} & {0.41/0.34} & {0.3/0.25} & {0.08/0.07} & {0.16/0.13} & {98.2} & {1.8} & {0.0} \\
&& {0.6} & {0.029} & {0.09} & {0.0} & {9.6/24.0} & {15.3/15.3} & {28.8/23.4} & {17.2/14.0} & {15.3/12.4} & {1.4/1.14} & {1.56/1.27} & {1.7/1.38} & {0.74/0.59} & {0.41/0.33} & {0.3/0.24} & {0.08/0.07} & {0.16/0.13} & {98.2} & {1.8} & {0.0} \\
&& {0.3} & {0.038} & {0.1} & {0.0} & {3.8/9.5} & {24.6/24.6} & {28.2/26.0} & {16.8/15.5} & {15.0/13.9} & {1.48/1.38} & {1.66/1.54} & {1.63/1.5} & {0.63/0.57} & {0.4/0.37} & {0.27/0.25} & {0.08/0.08} & {0.16/0.14} & {95.4} & {4.6} & {0.0} \\
&& {0.7} & {0.046} & {0.04} & {0.0} & {8.9/22.3} & {16.7/16.6} & {28.8/23.8} & {17.2/14.2} & {15.3/12.7} & {1.39/1.15} & {1.55/1.28} & {1.69/1.39} & {0.71/0.57} & {0.41/0.34} & {0.29/0.24} & {0.08/0.07} & {0.16/0.13} & {98.3} & {1.7} & {0.0} \\
&& {0.8} & {0.073} & {0.03} & {0.0} & {6.4/16.1} & {19.5/19.5} & {28.1/24.5} & {16.9/14.7} & {14.9/13.0} & {1.35/1.17} & {1.5/1.31} & {1.69/1.48} & {0.82/0.72} & {0.4/0.35} & {0.31/0.27} & {0.08/0.07} & {0.16/0.14} & {98.7} & {1.3} & {0.0} \\
&& {0.9} & {0.089} & {0.03} & {0.0} & {7.5/18.8} & {16.5/16.5} & {28.3/24.1} & {17.0/14.5} & {15.0/12.7} & {1.36/1.16} & {1.52/1.29} & {1.7/1.45} & {0.82/0.7} & {0.4/0.34} & {0.31/0.26} & {0.08/0.07} & {0.16/0.13} & {98.8} & {1.2} & {0.0} \\
\hline
\multirow{7}{*}{Run51} & \multirow{7}{*}{7} 
& {0.6} & {0.011} & {0.16} & {2.56} & {13.5/33.8} & {17.8/14.7} & {25.9/19.5} & {15.5/11.7} & {13.7/10.3} & {1.25/0.95} & {1.4/1.05} & {1.55/1.17} & {0.72/0.54} & {0.37/0.28} & {0.28/0.21} & {0.08/0.06} & {0.14/0.11} & {46.5} & {22.0} & {31.6} \\
&& {1.0} & {0.015} & {0.21} & {0.67} & {9.0/22.5} & {17.5/17.1} & {28.2/23.3} & {16.9/13.9} & {15.0/12.4} & {1.47/1.23} & {1.59/1.33} & {1.66/1.37} & {0.72/0.58} & {0.4/0.33} & {0.29/0.24} & {0.08/0.07} & {0.16/0.13} & {26.7} & {1.4} & {71.9} \\
&& {0.5} & {0.019} & {0.18} & {3.77} & {7.4/18.5} & {18.9/18.8} & {28.4/24.3} & {17.0/14.5} & {15.1/12.9} & {1.49/1.29} & {1.61/1.39} & {1.67/1.42} & {0.71/0.6} & {0.4/0.34} & {0.29/0.25} & {0.08/0.07} & {0.16/0.13} & {78.6} & {0.0} & {21.4} \\
&& {0.5} & {0.025} & {0.17} & {0.3} & {6.7/16.8} & {20.1/20.0} & {28.3/24.6} & {16.9/14.7} & {15.0/13.1} & {1.54/1.36} & {1.64/1.44} & {1.65/1.43} & {0.69/0.58} & {0.4/0.35} & {0.28/0.24} & {0.08/0.07} & {0.16/0.14} & {80.5} & {0.0} & {19.5} \\
&& {0.6} & {0.04} & {0.05} & {0.15} & {7.1/17.7} & {20.1/19.9} & {28.2/24.3} & {16.9/14.5} & {15.0/13.0} & {1.56/1.37} & {1.65/1.44} & {1.65/1.41} & {0.68/0.56} & {0.4/0.34} & {0.28/0.24} & {0.08/0.07} & {0.16/0.13} & {80.7} & {0.0} & {19.3} \\
&& {0.8} & {0.064} & {0.01} & {0.03} & {5.6/14.1} & {22.5/22.1} & {27.8/24.8} & {16.7/14.9} & {14.7/13.1} & {1.32/1.18} & {1.47/1.31} & {1.68/1.5} & {0.81/0.72} & {0.4/0.36} & {0.31/0.27} & {0.08/0.07} & {0.15/0.14} & {81.3} & {0.0} & {18.7} \\
&& {0.2} & {0.102} & {0.01} & {0.01} & {19.7/49.3} & {0.7/0.6} & {29.8/18.7} & {17.9/11.3} & {15.8/9.9} & {1.44/0.91} & {1.6/1.01} & {1.79/1.12} & {0.85/0.53} & {0.43/0.27} & {0.32/0.2} & {0.09/0.05} & {0.16/0.1} & {94.0} & {6.0} & {0.0} \\
\hline
\multirow{8}{*}{Run54} & \multirow{8}{*}{8} 
& {1.1} & {0.011} & {0.13} & {0.0} & {9.0/22.6} & {19.1/18.1} & {27.6/23.0} & {16.5/13.7} & {14.7/12.2} & {1.45/1.22} & {1.57/1.32} & {1.62/1.34} & {0.69/0.56} & {0.39/0.32} & {0.28/0.23} & {0.08/0.07} & {0.15/0.13} & {6.6} & {93.4} & {0.0} \\
&& {1.9} & {0.016} & {0.15} & {0.0} & {9.3/23.3} & {17.3/16.7} & {28.1/23.1} & {16.8/13.8} & {14.9/12.3} & {1.43/1.19} & {1.57/1.3} & {1.66/1.36} & {0.73/0.59} & {0.4/0.33} & {0.29/0.24} & {0.08/0.07} & {0.16/0.13} & {14.7} & {35.2} & {50.1} \\
&& {0.6} & {0.021} & {0.14} & {0.0} & {7.8/19.5} & {19.6/19.1} & {28.0/23.8} & {16.7/14.2} & {14.8/12.6} & {1.46/1.26} & {1.58/1.35} & {1.65/1.4} & {0.72/0.6} & {0.4/0.34} & {0.29/0.24} & {0.08/0.07} & {0.15/0.13} & {78.0} & {0.0} & {22.0} \\
&& {0.7} & {0.033} & {0.06} & {0.0} & {7.1/17.7} & {19.0/19.0} & {28.4/24.4} & {17.0/14.7} & {15.0/12.9} & {1.36/1.16} & {1.51/1.3} & {1.71/1.47} & {0.82/0.7} & {0.41/0.35} & {0.31/0.27} & {0.08/0.07} & {0.16/0.14} & {98.5} & {1.5} & {0.0} \\
&& {0.6} & {0.043} & {0.06} & {0.0} & {8.0/20.1} & {17.2/17.2} & {28.4/23.8} & {17.0/14.3} & {15.0/12.6} & {1.36/1.14} & {1.52/1.27} & {1.7/1.43} & {0.8/0.67} & {0.41/0.34} & {0.31/0.26} & {0.08/0.07} & {0.16/0.13} & {98.2} & {1.8} & {0.0} \\
&& {0.7} & {0.057} & {0.05} & {0.0} & {8.2/20.5} & {16.6/16.6} & {28.5/23.9} & {17.1/14.3} & {15.1/12.6} & {1.37/1.14} & {1.52/1.27} & {1.72/1.44} & {0.78/0.65} & {0.41/0.34} & {0.31/0.26} & {0.08/0.07} & {0.16/0.13} & {98.4} & {1.6} & {0.0} \\
&& {0.8} & {0.069} & {0.04} & {0.0} & {13.4/33.6} & {9.6/9.6} & {29.1/21.5} & {17.5/12.9} & {15.4/11.4} & {1.4/1.04} & {1.56/1.15} & {1.75/1.3} & {0.84/0.62} & {0.42/0.31} & {0.32/0.23} & {0.08/0.06} & {0.16/0.12} & {69.3} & {30.7} & {0.0} \\
&& {0.9} & {0.09} & {0.01} & {0.0} & {6.9/17.3} & {17.3/17.2} & {28.2/24.3} & {16.9/14.6} & {14.9/12.9} & {1.36/1.17} & {1.51/1.3} & {1.7/1.46} & {0.82/0.71} & {0.4/0.35} & {0.31/0.26} & {0.08/0.07} & {0.16/0.13} & {98.9} & {1.1} & {0.0} \\
\hline
\multirow{7}{*}{Run55} & \multirow{7}{*}{7} 
& {0.8} & {0.012} & {0.08} & {0.0} & {13.3/33.2} & {17.1/14.4} & {26.3/19.9} & {15.8/11.9} & {13.9/10.5} & {1.27/0.96} & {1.41/1.07} & {1.57/1.19} & {0.73/0.55} & {0.38/0.28} & {0.28/0.21} & {0.08/0.06} & {0.15/0.11} & {14.5} & {1.5} & {84.0} \\
&& {1.0} & {0.016} & {0.1} & {0.0} & {7.3/18.2} & {19.4/19.2} & {28.2/24.2} & {16.9/14.5} & {15.0/12.9} & {1.49/1.3} & {1.6/1.39} & {1.66/1.42} & {0.71/0.6} & {0.4/0.34} & {0.29/0.25} & {0.08/0.07} & {0.16/0.13} & {40.3} & {0.0} & {59.7} \\
&& {0.5} & {0.025} & {0.07} & {0.0} & {5.4/13.4} & {24.2/23.5} & {27.6/24.9} & {16.5/14.8} & {14.7/13.3} & {1.61/1.48} & {1.67/1.52} & {1.59/1.43} & {0.61/0.53} & {0.39/0.35} & {0.27/0.24} & {0.08/0.07} & {0.15/0.14} & {80.7} & {0.0} & {19.3} \\
&& {0.6} & {0.028} & {0.05} & {0.0} & {7.2/18.1} & {20.0/19.8} & {28.2/24.3} & {16.9/14.5} & {15.0/12.9} & {1.58/1.39} & {1.66/1.45} & {1.64/1.4} & {0.66/0.55} & {0.4/0.34} & {0.28/0.24} & {0.08/0.07} & {0.16/0.13} & {82.5} & {0.0} & {17.5} \\
&& {0.6} & {0.032} & {0.05} & {0.0} & {13.2/32.9} & {10.2/10.1} & {29.0/21.7} & {17.4/13.0} & {15.4/11.5} & {1.41/1.05} & {1.57/1.17} & {1.73/1.29} & {0.8/0.59} & {0.41/0.31} & {0.31/0.23} & {0.08/0.06} & {0.16/0.12} & {52.9} & {0.0} & {47.1} \\
&& {0.7} & {0.052} & {0.02} & {0.0} & {6.1/15.2} & {21.0/21.0} & {28.5/25.1} & {17.1/15.0} & {15.1/13.3} & {1.35/1.19} & {1.51/1.33} & {1.68/1.48} & {0.68/0.59} & {0.41/0.36} & {0.29/0.25} & {0.08/0.07} & {0.16/0.14} & {98.3} & {1.7} & {0.0} \\
&& {0.8} & {0.068} & {0.01} & {0.0} & {5.7/14.3} & {21.1/21.1} & {28.2/25.0} & {16.9/15.0} & {14.9/13.2} & {1.34/1.19} & {1.5/1.33} & {1.7/1.51} & {0.83/0.74} & {0.4/0.36} & {0.31/0.27} & {0.08/0.07} & {0.16/0.14} & {98.5} & {1.5} & {0.0} \\
\hline
\multirow{8}{*}{Run60} & \multirow{8}{*}{8} 
& {0.4} & {0.012} & {0.18} & {0.0} & {6.8/17.0} & {25.3/23.3} & {26.4/23.3} & {15.8/13.9} & {14.0/12.4} & {1.46/1.31} & {1.54/1.38} & {1.54/1.35} & {0.63/0.54} & {0.37/0.33} & {0.26/0.23} & {0.08/0.07} & {0.15/0.13} & {41.9} & {0.0} & {58.1} \\
&& {1.2} & {0.016} & {0.17} & {0.0} & {11.3/28.3} & {17.3/15.6} & {27.1/21.5} & {16.3/12.9} & {14.4/11.4} & {1.35/1.07} & {1.49/1.18} & {1.61/1.27} & {0.73/0.56} & {0.39/0.31} & {0.29/0.22} & {0.08/0.06} & {0.15/0.12} & {17.0} & {4.9} & {78.1} \\
&& {0.5} & {0.025} & {0.18} & {0.0} & {5.5/13.8} & {21.8/21.8} & {28.2/25.2} & {16.9/15.0} & {15.0/13.4} & {1.55/1.4} & {1.65/1.48} & {1.65/1.46} & {0.67/0.58} & {0.4/0.35} & {0.28/0.25} & {0.08/0.07} & {0.16/0.14} & {97.8} & {2.2} & {0.0} \\
&& {0.5} & {0.032} & {0.08} & {0.0} & {7.6/18.9} & {18.7/18.7} & {28.5/24.2} & {17.0/14.5} & {15.1/12.9} & {1.51/1.3} & {1.63/1.4} & {1.67/1.42} & {0.7/0.58} & {0.4/0.34} & {0.29/0.24} & {0.08/0.07} & {0.16/0.13} & {97.9} & {2.1} & {0.0} \\
&& {0.7} & {0.052} & {0.05} & {0.0} & {5.9/14.7} & {21.5/21.5} & {28.5/25.2} & {17.0/15.1} & {15.1/13.4} & {1.35/1.19} & {1.51/1.33} & {1.67/1.48} & {0.66/0.56} & {0.4/0.36} & {0.28/0.25} & {0.08/0.07} & {0.16/0.14} & {85.2} & {0.0} & {14.8} \\
&& {0.9} & {0.082} & {0.02} & {0.0} & {5.0/12.4} & {20.6/20.6} & {27.7/24.9} & {16.6/14.9} & {14.6/13.2} & {1.33/1.19} & {1.48/1.33} & {1.67/1.5} & {0.81/0.73} & {0.4/0.36} & {0.3/0.27} & {0.08/0.07} & {0.15/0.14} & {88.7} & {0.0} & {11.3} \\
&& {0.2} & {0.107} & {0.04} & {0.0} & {20.0/50.0} & {0.2/0.1} & {29.8/18.6} & {17.9/11.2} & {15.8/9.9} & {1.44/0.9} & {1.6/1.0} & {1.79/1.12} & {0.85/0.53} & {0.43/0.27} & {0.32/0.2} & {0.09/0.05} & {0.17/0.1} & {95.2} & {4.8} & {0.0} \\
&& {0.2} & {0.13} & {0.01} & {0.0} & {20.0/50.0} & {0.2/0.1} & {29.8/18.6} & {17.9/11.2} & {15.8/9.9} & {1.44/0.9} & {1.6/1.0} & {1.79/1.12} & {0.85/0.53} & {0.43/0.27} & {0.32/0.2} & {0.09/0.05} & {0.17/0.1} & {95.6} & {4.4} & {0.0} \\
\hline
\multirow{8}{*}{Run68} & \multirow{8}{*}{8} 
& {1.0} & {0.011} & {0.11} & {0.0} & {12.2/30.6} & {17.3/15.0} & {26.5/20.6} & {15.9/12.3} & {14.0/10.9} & {1.28/1.0} & {1.43/1.11} & {1.59/1.23} & {0.75/0.58} & {0.38/0.29} & {0.29/0.22} & {0.08/0.06} & {0.15/0.11} & {27.7} & {7.3} & {65.0} \\
&& {1.0} & {0.015} & {0.13} & {0.0} & {6.1/15.2} & {21.5/21.2} & {28.0/24.7} & {16.7/14.8} & {14.9/13.1} & {1.51/1.35} & {1.61/1.44} & {1.64/1.44} & {0.69/0.59} & {0.4/0.35} & {0.28/0.25} & {0.08/0.07} & {0.15/0.14} & {40.6} & {1.2} & {58.2} \\
&& {0.5} & {0.019} & {0.11} & {0.0} & {7.6/19.1} & {18.5/18.5} & {28.5/24.2} & {17.0/14.5} & {15.1/12.9} & {1.5/1.29} & {1.63/1.4} & {1.67/1.42} & {0.71/0.59} & {0.4/0.34} & {0.29/0.24} & {0.08/0.07} & {0.16/0.13} & {80.7} & {0.0} & {19.3} \\
&& {0.6} & {0.025} & {0.05} & {0.0} & {11.3/28.2} & {13.1/13.1} & {28.8/22.5} & {17.2/13.5} & {15.3/11.9} & {1.46/1.16} & {1.6/1.26} & {1.71/1.33} & {0.76/0.58} & {0.41/0.32} & {0.3/0.23} & {0.08/0.07} & {0.16/0.12} & {59.7} & {0.0} & {40.3} \\
&& {0.5} & {0.033} & {0.05} & {0.0} & {9.1/22.8} & {16.7/16.5} & {28.6/23.5} & {17.1/14.1} & {15.2/12.5} & {1.48/1.23} & {1.61/1.33} & {1.68/1.38} & {0.73/0.58} & {0.41/0.33} & {0.29/0.24} & {0.08/0.07} & {0.16/0.13} & {81.7} & {0.0} & {18.3} \\
&& {0.6} & {0.044} & {0.05} & {0.0} & {8.3/20.9} & {17.7/17.6} & {28.7/24.0} & {17.2/14.4} & {15.2/12.7} & {1.38/1.15} & {1.54/1.29} & {1.7/1.42} & {0.71/0.58} & {0.41/0.34} & {0.29/0.24} & {0.08/0.07} & {0.16/0.13} & {98.3} & {1.7} & {0.0} \\
&& {0.7} & {0.057} & {0.04} & {0.0} & {8.3/20.9} & {17.6/17.6} & {28.6/23.9} & {17.2/14.4} & {15.1/12.7} & {1.36/1.14} & {1.52/1.27} & {1.72/1.44} & {0.75/0.62} & {0.41/0.34} & {0.31/0.26} & {0.08/0.07} & {0.16/0.13} & {98.6} & {1.4} & {0.0} \\
&& {0.8} & {0.075} & {0.01} & {0.0} & {7.0/17.5} & {18.8/18.5} & {27.8/24.0} & {16.7/14.4} & {14.7/12.7} & {1.34/1.15} & {1.49/1.28} & {1.67/1.45} & {0.81/0.7} & {0.4/0.34} & {0.3/0.26} & {0.08/0.07} & {0.15/0.13} & {88.2} & {0.0} & {11.8} \\
\hline
\multirow{6}{*}{Run73} & \multirow{6}{*}{6} 
& {1.0} & {0.011} & {0.06} & {0.0} & {12.3/30.7} & {16.7/14.7} & {26.8/20.8} & {16.1/12.5} & {14.2/11.0} & {1.3/1.01} & {1.45/1.12} & {1.61/1.24} & {0.74/0.57} & {0.38/0.3} & {0.29/0.22} & {0.08/0.06} & {0.15/0.12} & {66.5} & {15.3} & {18.2} \\
&& {0.7} & {0.015} & {0.12} & {0.0} & {10.2/25.4} & {16.6/15.8} & {27.9/22.5} & {16.7/13.5} & {14.8/12.0} & {1.39/1.13} & {1.53/1.24} & {1.66/1.34} & {0.75/0.59} & {0.4/0.32} & {0.29/0.24} & {0.08/0.07} & {0.15/0.12} & {34.3} & {21.4} & {44.3} \\
&& {0.7} & {0.019} & {0.05} & {0.0} & {5.8/14.5} & {21.5/21.4} & {28.2/25.0} & {16.8/14.9} & {15.0/13.3} & {1.52/1.37} & {1.63/1.46} & {1.65/1.45} & {0.68/0.59} & {0.4/0.35} & {0.28/0.25} & {0.08/0.07} & {0.16/0.14} & {52.7} & {33.6} & {13.7} \\
&& {0.6} & {0.025} & {0.05} & {0.0} & {7.2/18.1} & {19.0/19.0} & {28.4/24.4} & {17.0/14.6} & {15.1/13.0} & {1.46/1.26} & {1.6/1.38} & {1.67/1.43} & {0.71/0.6} & {0.4/0.35} & {0.29/0.25} & {0.08/0.07} & {0.16/0.13} & {82.2} & {0.0} & {17.8} \\
&& {0.5} & {0.033} & {0.04} & {0.0} & {7.8/19.5} & {18.3/18.3} & {28.5/24.2} & {17.1/14.4} & {15.2/12.8} & {1.5/1.29} & {1.63/1.4} & {1.67/1.41} & {0.71/0.58} & {0.4/0.34} & {0.29/0.24} & {0.08/0.07} & {0.16/0.13} & {81.5} & {0.0} & {18.5} \\
&& {0.7} & {0.044} & {0.02} & {0.0} & {7.0/17.4} & {19.8/19.7} & {28.5/24.6} & {17.1/14.8} & {15.1/13.1} & {1.36/1.17} & {1.51/1.3} & {1.7/1.47} & {0.7/0.59} & {0.41/0.35} & {0.29/0.25} & {0.08/0.07} & {0.16/0.14} & {98.5} & {1.5} & {0.0} \\
\hline
\multirow{6}{*}{Run74} & \multirow{6}{*}{6} 
& {1.5} & {0.013} & {0.1} & {0.0} & {10.0/25.0} & {19.4/17.7} & {27.0/22.1} & {16.2/13.2} & {14.3/11.7} & {1.37/1.14} & {1.5/1.24} & {1.6/1.3} & {0.7/0.56} & {0.38/0.31} & {0.28/0.23} & {0.08/0.06} & {0.15/0.12} & {13.2} & {3.5} & {83.3} \\
&& {0.7} & {0.016} & {0.18} & {0.0} & {4.6/11.6} & {22.9/22.9} & {28.1/25.5} & {16.8/15.2} & {14.9/13.6} & {1.49/1.37} & {1.6/1.46} & {1.65/1.49} & {0.69/0.61} & {0.4/0.36} & {0.28/0.26} & {0.08/0.08} & {0.16/0.14} & {54.5} & {32.0} & {13.5} \\
&& {0.5} & {0.02} & {0.17} & {0.0} & {6.5/16.3} & {22.6/21.7} & {27.5/24.2} & {16.4/14.4} & {14.6/12.9} & {1.46/1.3} & {1.58/1.4} & {1.6/1.41} & {0.66/0.57} & {0.39/0.34} & {0.28/0.24} & {0.08/0.07} & {0.15/0.13} & {76.7} & {3.9} & {19.4} \\
&& {0.6} & {0.026} & {0.08} & {0.0} & {7.1/17.8} & {19.5/19.4} & {28.5/24.5} & {17.0/14.7} & {15.1/13.0} & {1.42/1.22} & {1.56/1.35} & {1.68/1.44} & {0.69/0.58} & {0.4/0.35} & {0.29/0.24} & {0.08/0.07} & {0.16/0.14} & {98.2} & {1.8} & {0.0} \\
&& {0.6} & {0.041} & {0.04} & {0.0} & {6.8/17.0} & {19.8/19.8} & {28.5/24.7} & {17.1/14.8} & {15.1/13.1} & {1.41/1.23} & {1.56/1.35} & {1.69/1.46} & {0.7/0.59} & {0.41/0.35} & {0.29/0.25} & {0.08/0.07} & {0.16/0.14} & {81.4} & {0.0} & {18.6} \\
&& {0.8} & {0.054} & {0.02} & {0.0} & {6.0/14.9} & {21.1/20.9} & {27.9/24.7} & {16.8/14.8} & {14.8/13.1} & {1.34/1.18} & {1.49/1.31} & {1.69/1.49} & {0.82/0.73} & {0.4/0.35} & {0.31/0.27} & {0.08/0.07} & {0.15/0.14} & {87.2} & {0.0} & {12.8} \\
\hline
\multirow{8}{*}{Run81} & \multirow{8}{*}{8} 
& {0.9} & {0.011} & {0.18} & {0.0} & {9.0/22.6} & {20.6/19.0} & {27.1/22.6} & {16.2/13.5} & {14.4/12.0} & {1.42/1.2} & {1.53/1.29} & {1.59/1.33} & {0.69/0.56} & {0.38/0.32} & {0.28/0.23} & {0.08/0.07} & {0.15/0.12} & {80.6} & {6.5} & {12.9} \\
&& {1.6} & {0.018} & {0.06} & {0.0} & {11.7/29.1} & {14.0/13.4} & {28.3/22.0} & {16.9/13.2} & {15.0/11.7} & {1.42/1.11} & {1.56/1.22} & {1.68/1.3} & {0.76/0.58} & {0.4/0.31} & {0.3/0.23} & {0.08/0.06} & {0.16/0.12} & {43.5} & {15.2} & {41.3} \\
&& {0.5} & {0.023} & {0.14} & {0.0} & {8.0/19.9} & {17.9/17.9} & {28.5/24.1} & {17.1/14.4} & {15.1/12.8} & {1.49/1.28} & {1.61/1.37} & {1.68/1.41} & {0.72/0.59} & {0.4/0.34} & {0.29/0.24} & {0.08/0.07} & {0.16/0.13} & {97.1} & {2.9} & {0.0} \\
&& {0.6} & {0.031} & {0.05} & {0.0} & {9.8/24.5} & {15.4/15.4} & {28.7/23.2} & {17.2/13.9} & {15.2/12.3} & {1.5/1.23} & {1.62/1.32} & {1.69/1.36} & {0.73/0.58} & {0.41/0.33} & {0.29/0.24} & {0.08/0.07} & {0.16/0.13} & {98.1} & {1.9} & {0.0} \\
&& {0.6} & {0.04} & {0.05} & {0.0} & {10.9/27.3} & {13.7/13.6} & {28.9/22.8} & {17.3/13.6} & {15.3/12.1} & {1.44/1.14} & {1.59/1.26} & {1.71/1.34} & {0.75/0.58} & {0.41/0.32} & {0.3/0.23} & {0.08/0.07} & {0.16/0.13} & {98.2} & {1.8} & {0.0} \\
&& {0.8} & {0.064} & {0.01} & {0.0} & {6.8/17.0} & {19.8/19.7} & {28.3/24.5} & {17.0/14.7} & {15.0/13.0} & {1.35/1.17} & {1.51/1.3} & {1.71/1.48} & {0.82/0.71} & {0.41/0.35} & {0.31/0.27} & {0.08/0.07} & {0.16/0.14} & {98.6} & {1.4} & {0.0} \\
&& {0.3} & {0.102} & {0.02} & {0.0} & {20.0/49.9} & {0.5/0.3} & {29.8/18.7} & {17.9/11.2} & {15.8/9.9} & {1.45/0.9} & {1.61/1.0} & {1.79/1.12} & {0.84/0.53} & {0.43/0.27} & {0.32/0.2} & {0.09/0.05} & {0.17/0.1} & {96.4} & {3.6} & {0.0} \\
&& {0.4} & {0.133} & {0.02} & {0.0} & {20.0/50.0} & {0.1/0.0} & {30.0/18.8} & {18.0/11.3} & {15.9/9.9} & {1.45/0.91} & {1.61/1.01} & {1.8/1.12} & {0.85/0.53} & {0.43/0.27} & {0.32/0.2} & {0.09/0.05} & {0.17/0.1} & {97.2} & {2.8} & {0.0} \\
\hline
\multirow{9}{*}{Run93} & \multirow{9}{*}{9} 
& {1.8} & {0.012} & {0.1} & {0.0} & {13.1/32.8} & {12.9/11.8} & {27.8/20.9} & {16.7/12.6} & {14.7/11.1} & {1.35/1.01} & {1.49/1.12} & {1.67/1.26} & {0.79/0.59} & {0.4/0.3} & {0.3/0.23} & {0.08/0.06} & {0.15/0.12} & {49.7} & {39.0} & {11.3} \\
&& {1.1} & {0.017} & {0.2} & {0.0} & {10.1/25.3} & {15.4/15.1} & {28.5/22.9} & {17.1/13.7} & {15.1/12.2} & {1.44/1.17} & {1.58/1.28} & {1.69/1.35} & {0.74/0.58} & {0.41/0.33} & {0.3/0.24} & {0.08/0.07} & {0.16/0.13} & {32.1} & {40.9} & {27.0} \\
&& {0.5} & {0.023} & {0.07} & {0.0} & {9.4/23.5} & {16.0/15.9} & {28.6/23.4} & {17.1/14.0} & {15.2/12.4} & {1.48/1.23} & {1.61/1.33} & {1.69/1.37} & {0.73/0.58} & {0.41/0.33} & {0.29/0.24} & {0.08/0.07} & {0.16/0.13} & {80.8} & {0.0} & {19.2} \\
&& {0.3} & {0.03} & {0.12} & {0.0} & {1.4/3.4} & {29.5/29.5} & {28.0/27.3} & {16.6/16.1} & {15.0/14.6} & {2.01/1.97} & {1.94/1.9} & {1.54/1.49} & {0.45/0.43} & {0.39/0.37} & {0.24/0.23} & {0.08/0.08} & {0.15/0.15} & {75.8} & {24.2} & {0.0} \\
&& {0.6} & {0.036} & {0.09} & {0.0} & {9.5/23.8} & {15.7/15.6} & {28.7/23.4} & {17.2/14.0} & {15.2/12.4} & {1.45/1.19} & {1.59/1.3} & {1.7/1.38} & {0.74/0.59} & {0.41/0.33} & {0.3/0.24} & {0.08/0.07} & {0.16/0.13} & {82.1} & {0.0} & {17.9} \\
&& {0.6} & {0.048} & {0.04} & {0.0} & {9.1/22.8} & {16.5/16.5} & {28.8/23.7} & {17.2/14.1} & {15.3/12.6} & {1.43/1.19} & {1.6/1.33} & {1.68/1.38} & {0.71/0.56} & {0.41/0.33} & {0.29/0.24} & {0.08/0.07} & {0.16/0.13} & {98.3} & {1.7} & {0.0} \\
&& {0.8} & {0.076} & {0.01} & {0.0} & {7.2/18.0} & {18.2/18.2} & {28.1/24.1} & {16.9/14.5} & {14.9/12.7} & {1.35/1.15} & {1.5/1.28} & {1.69/1.45} & {0.82/0.71} & {0.4/0.34} & {0.31/0.26} & {0.08/0.07} & {0.16/0.13} & {98.7} & {1.3} & {0.0} \\
&& {0.3} & {0.12} & {0.01} & {0.0} & {19.9/49.7} & {0.6/0.4} & {29.8/18.7} & {17.9/11.2} & {15.8/9.9} & {1.45/0.91} & {1.61/1.01} & {1.79/1.12} & {0.84/0.53} & {0.43/0.27} & {0.32/0.2} & {0.09/0.05} & {0.17/0.1} & {96.3} & {3.7} & {0.0} \\
&& {0.3} & {0.191} & {0.0} & {0.0} & {19.7/49.2} & {0.6/0.5} & {29.9/18.9} & {18.0/11.3} & {15.9/10.0} & {1.45/0.91} & {1.61/1.02} & {1.79/1.13} & {0.85/0.53} & {0.43/0.27} & {0.32/0.2} & {0.09/0.06} & {0.17/0.1} & {96.2} & {3.8} & {0.0} \\
\hline
\multirow{6}{*}{Run94} & \multirow{6}{*}{6} 
& {1.7} & {0.012} & {0.03} & {0.0} & {10.8/27.0} & {17.4/16.0} & {27.4/21.9} & {16.4/13.1} & {14.6/11.6} & {1.42/1.16} & {1.54/1.24} & {1.62/1.29} & {0.71/0.55} & {0.39/0.31} & {0.28/0.22} & {0.08/0.06} & {0.15/0.12} & {21.3} & {2.5} & {76.2} \\
&& {1.0} & {0.017} & {0.1} & {0.0} & {9.9/24.8} & {15.0/15.0} & {28.7/23.2} & {17.2/13.9} & {15.2/12.3} & {1.47/1.2} & {1.6/1.3} & {1.7/1.37} & {0.75/0.59} & {0.41/0.33} & {0.3/0.24} & {0.08/0.07} & {0.16/0.13} & {45.4} & {28.3} & {26.3} \\
&& {0.6} & {0.023} & {0.06} & {0.0} & {13.3/33.4} & {11.3/10.7} & {28.4/21.2} & {17.1/12.7} & {15.1/11.2} & {1.37/1.02} & {1.53/1.14} & {1.7/1.27} & {0.79/0.58} & {0.41/0.3} & {0.3/0.23} & {0.08/0.06} & {0.16/0.12} & {64.5} & {2.1} & {33.4} \\
&& {0.7} & {0.03} & {0.03} & {0.0} & {6.7/16.8} & {19.4/19.4} & {28.2/24.5} & {17.0/14.7} & {15.0/13.0} & {1.35/1.17} & {1.5/1.3} & {1.7/1.47} & {0.82/0.71} & {0.4/0.35} & {0.31/0.27} & {0.08/0.07} & {0.16/0.14} & {97.5} & {2.5} & {0.0} \\
&& {0.5} & {0.042} & {0.03} & {0.0} & {3.0/7.6} & {24.0/23.9} & {27.4/25.8} & {16.5/15.5} & {14.5/13.6} & {1.31/1.23} & {1.46/1.37} & {1.65/1.55} & {0.8/0.75} & {0.39/0.37} & {0.3/0.28} & {0.08/0.08} & {0.15/0.14} & {97.7} & {2.3} & {0.0} \\
&& {0.8} & {0.048} & {0.02} & {0.0} & {6.8/17.1} & {17.7/17.6} & {28.2/24.4} & {16.9/14.6} & {14.9/12.9} & {1.36/1.17} & {1.51/1.3} & {1.69/1.47} & {0.82/0.71} & {0.4/0.35} & {0.31/0.27} & {0.08/0.07} & {0.16/0.14} & {98.5} & {1.5} & {0.0} \\
\hline
\multirow{8}{*}{Run95} & \multirow{8}{*}{8} 
& {0.5} & {0.013} & {0.19} & {0.0} & {7.6/19.1} & {24.8/22.5} & {26.2/22.8} & {15.7/13.6} & {13.9/12.1} & {1.36/1.2} & {1.48/1.3} & {1.54/1.34} & {0.65/0.55} & {0.37/0.32} & {0.27/0.23} & {0.08/0.07} & {0.15/0.13} & {41.7} & {16.2} & {42.2} \\
&& {1.2} & {0.016} & {0.12} & {0.0} & {13.1/32.8} & {11.7/11.0} & {28.3/21.2} & {17.0/12.7} & {15.0/11.2} & {1.37/1.02} & {1.52/1.14} & {1.7/1.27} & {0.8/0.6} & {0.4/0.3} & {0.31/0.23} & {0.08/0.06} & {0.16/0.12} & {31.6} & {11.0} & {57.4} \\
&& {0.9} & {0.02} & {0.14} & {0.0} & {12.1/30.3} & {13.5/12.8} & {28.2/21.7} & {16.9/13.0} & {14.9/11.5} & {1.38/1.06} & {1.53/1.18} & {1.68/1.29} & {0.77/0.59} & {0.4/0.31} & {0.3/0.23} & {0.08/0.06} & {0.16/0.12} & {58.5} & {12.6} & {28.9} \\
&& {0.5} & {0.032} & {0.08} & {0.0} & {10.1/25.3} & {15.9/15.4} & {28.2/22.8} & {16.9/13.6} & {15.0/12.1} & {1.43/1.17} & {1.57/1.27} & {1.67/1.35} & {0.75/0.59} & {0.4/0.32} & {0.3/0.24} & {0.08/0.07} & {0.16/0.13} & {39.7} & {41.9} & {18.4} \\
&& {0.8} & {0.041} & {0.05} & {0.0} & {17.9/44.7} & {4.4/3.8} & {29.0/19.2} & {17.4/11.5} & {15.4/10.2} & {1.4/0.93} & {1.56/1.03} & {1.74/1.15} & {0.83/0.55} & {0.41/0.27} & {0.31/0.21} & {0.08/0.06} & {0.16/0.11} & {40.2} & {2.9} & {56.9} \\
&& {0.8} & {0.066} & {0.04} & {0.0} & {5.6/14.0} & {21.9/21.6} & {27.9/24.9} & {16.8/14.9} & {14.8/13.2} & {1.33/1.18} & {1.48/1.32} & {1.68/1.5} & {0.82/0.73} & {0.4/0.36} & {0.31/0.27} & {0.08/0.07} & {0.15/0.14} & {85.9} & {0.0} & {14.1} \\
&& {0.9} & {0.076} & {0.02} & {0.0} & {7.8/19.6} & {17.3/17.3} & {28.2/23.8} & {16.9/14.3} & {14.9/12.6} & {1.36/1.14} & {1.51/1.27} & {1.7/1.43} & {0.82/0.7} & {0.4/0.34} & {0.31/0.26} & {0.08/0.07} & {0.16/0.13} & {98.6} & {1.4} & {0.0} \\
&& {0.3} & {0.121} & {0.0} & {0.0} & {20.0/50.0} & {0.0/0.0} & {30.0/18.7} & {18.0/11.2} & {15.9/9.9} & {1.45/0.91} & {1.61/1.01} & {1.8/1.12} & {0.85/0.53} & {0.43/0.27} & {0.32/0.2} & {0.09/0.05} & {0.17/0.1} & {96.4} & {3.6} & {0.0} \\
\hline
\multirow{9}{*}{Run98} & \multirow{9}{*}{9} 
& {0.7} & {0.01} & {0.11} & {0.0} & {14.8/37.1} & {13.4/11.2} & {26.9/19.5} & {16.2/11.7} & {14.3/10.3} & {1.3/0.94} & {1.44/1.04} & {1.62/1.17} & {0.77/0.55} & {0.39/0.28} & {0.29/0.21} & {0.08/0.06} & {0.15/0.11} & {25.9} & {18.9} & {55.2} \\
&& {1.6} & {0.016} & {0.04} & {0.0} & {13.5/33.7} & {10.8/10.3} & {28.6/21.2} & {17.1/12.7} & {15.1/11.2} & {1.39/1.03} & {1.54/1.15} & {1.71/1.27} & {0.79/0.58} & {0.41/0.3} & {0.31/0.23} & {0.08/0.06} & {0.16/0.12} & {22.0} & {33.0} & {45.0} \\
&& {0.7} & {0.021} & {0.09} & {0.0} & {9.0/22.5} & {18.0/17.3} & {28.0/23.2} & {16.7/13.9} & {14.8/12.3} & {1.41/1.18} & {1.55/1.29} & {1.66/1.37} & {0.73/0.59} & {0.4/0.33} & {0.29/0.24} & {0.08/0.07} & {0.15/0.13} & {52.0} & {25.3} & {22.7} \\
&& {0.6} & {0.028} & {0.04} & {0.0} & {8.7/21.7} & {17.3/17.2} & {28.7/23.8} & {17.1/14.2} & {15.2/12.7} & {1.47/1.23} & {1.64/1.38} & {1.68/1.39} & {0.7/0.56} & {0.41/0.34} & {0.29/0.24} & {0.08/0.07} & {0.16/0.13} & {98.3} & {1.7} & {0.0} \\
&& {0.6} & {0.037} & {0.03} & {0.0} & {9.9/24.8} & {15.1/15.1} & {28.8/23.2} & {17.3/13.9} & {15.3/12.3} & {1.45/1.18} & {1.61/1.31} & {1.7/1.37} & {0.74/0.58} & {0.41/0.33} & {0.3/0.24} & {0.08/0.07} & {0.16/0.13} & {98.2} & {1.8} & {0.0} \\
&& {0.7} & {0.048} & {0.03} & {0.0} & {9.9/24.7} & {15.2/15.2} & {28.8/23.2} & {17.3/13.9} & {15.3/12.3} & {1.39/1.12} & {1.55/1.25} & {1.72/1.39} & {0.76/0.6} & {0.41/0.33} & {0.31/0.25} & {0.08/0.07} & {0.16/0.13} & {97.8} & {2.2} & {0.0} \\
&& {0.8} & {0.064} & {0.01} & {0.0} & {8.8/21.9} & {16.3/16.3} & {28.5/23.6} & {17.1/14.2} & {15.1/12.5} & {1.37/1.13} & {1.52/1.26} & {1.72/1.42} & {0.82/0.68} & {0.41/0.34} & {0.31/0.26} & {0.08/0.07} & {0.16/0.13} & {98.7} & {1.3} & {0.0} \\
&& {0.3} & {0.101} & {0.02} & {0.0} & {20.0/49.9} & {0.1/0.1} & {30.0/18.8} & {18.0/11.3} & {15.9/9.9} & {1.45/0.91} & {1.61/1.01} & {1.8/1.12} & {0.85/0.53} & {0.43/0.27} & {0.32/0.2} & {0.09/0.05} & {0.17/0.1} & {97.0} & {3.0} & {0.0} \\
&& {0.4} & {0.132} & {0.02} & {0.0} & {20.0/49.9} & {0.1/0.1} & {30.0/18.8} & {18.0/11.3} & {15.9/9.9} & {1.45/0.91} & {1.61/1.01} & {1.8/1.12} & {0.85/0.53} & {0.43/0.27} & {0.32/0.2} & {0.09/0.05} & {0.17/0.1} & {97.2} & {2.8} & {0.0} \\
\hline
\multirow{6}{*}{Run100} & \multirow{6}{*}{6} 
& {0.4} & {0.013} & {0.11} & {0.0} & {9.9/24.7} & {19.6/17.8} & {26.9/22.0} & {16.1/13.2} & {14.2/11.7} & {1.33/1.09} & {1.47/1.21} & {1.6/1.31} & {0.72/0.58} & {0.38/0.31} & {0.28/0.23} & {0.08/0.06} & {0.15/0.12} & {29.4} & {6.2} & {64.4} \\
&& {1.2} & {0.016} & {0.1} & {0.0} & {12.0/30.1} & {14.1/13.2} & {28.0/21.6} & {16.8/13.0} & {14.8/11.5} & {1.4/1.09} & {1.54/1.19} & {1.67/1.28} & {0.76/0.58} & {0.4/0.31} & {0.3/0.23} & {0.08/0.06} & {0.16/0.12} & {23.7} & {1.4} & {74.9} \\
&& {0.4} & {0.022} & {0.14} & {0.0} & {3.2/8.1} & {25.1/25.1} & {28.0/26.2} & {16.7/15.6} & {14.9/13.9} & {1.53/1.45} & {1.63/1.53} & {1.63/1.52} & {0.66/0.61} & {0.4/0.37} & {0.28/0.26} & {0.08/0.08} & {0.15/0.14} & {97.0} & {3.0} & {0.0} \\
&& {0.5} & {0.026} & {0.05} & {0.0} & {7.3/18.2} & {18.8/18.8} & {28.4/24.4} & {17.0/14.6} & {15.1/12.9} & {1.48/1.28} & {1.6/1.38} & {1.68/1.43} & {0.72/0.6} & {0.4/0.35} & {0.29/0.25} & {0.08/0.07} & {0.16/0.13} & {80.6} & {0.0} & {19.4} \\
&& {0.5} & {0.034} & {0.05} & {0.0} & {7.0/17.5} & {19.7/19.6} & {28.4/24.5} & {17.0/14.6} & {15.1/13.0} & {1.54/1.35} & {1.64/1.43} & {1.66/1.42} & {0.68/0.57} & {0.4/0.35} & {0.28/0.24} & {0.08/0.07} & {0.16/0.14} & {97.6} & {2.4} & {0.0} \\
&& {0.7} & {0.054} & {0.0} & {0.0} & {6.1/15.2} & {21.1/21.1} & {28.5/25.1} & {17.1/15.1} & {15.1/13.3} & {1.35/1.18} & {1.51/1.32} & {1.7/1.5} & {0.68/0.58} & {0.41/0.36} & {0.29/0.26} & {0.08/0.07} & {0.16/0.14} & {98.5} & {1.5} & {0.0} 
\\
\hline
\end{longtable}

\endgroup
\end{landscape}

\clearpage
\twocolumn

\section{Re-scaling the dust condensates}\label{sec:appendix_B}

\begin{figure}
	\includegraphics[width=.95\columnwidth]{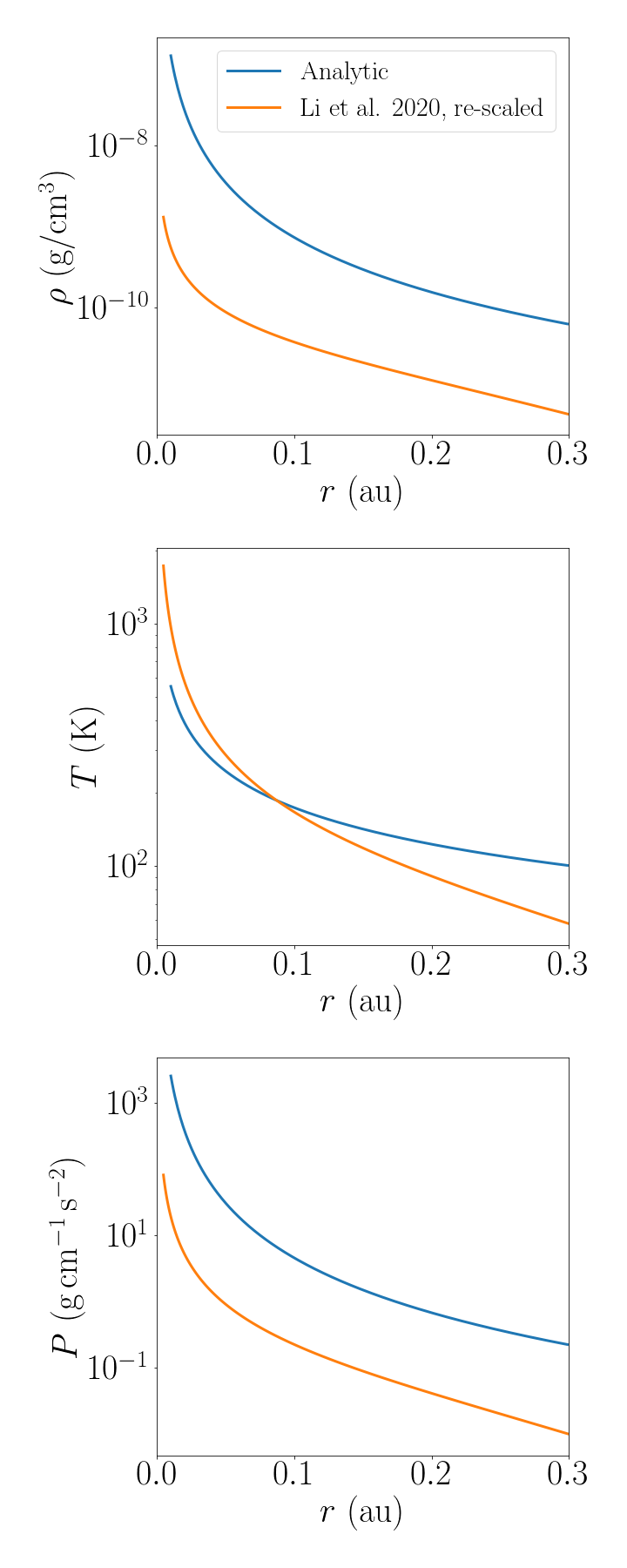}
    \caption{The density of the gas disc $\rho$, the temperature $T$ and mid-plane pressure $P$ profiles over 0.01--0.3\,au.  The orange data is re-scaled from the solar system disc used in the dust condensation simulation, to be consistent with the size of the disc and stellar mass of T1.  The blue data are the analytic profiles at $t=0$ used in the N-body simulations for the disc around T1.} 
    \label{fig:profiles}
\end{figure}

We re-scale the dust condensation model around a Sun-like star from \cite{Li2020} to a model for the dust condensation around a T1-like star.
To this end, we re-scale the size of the disc by matching the location of the ice line.  The ice line, the location where $T=170 \, \rm K$, is at $6.5 \, \rm au$ around a Sun-like star at the time we take from the \citet{Li2020} model.
We multiply the radii of the Sun-like disk by $0.1/6.5$ such that the new ice line resides at $0.1 \, \rm au$, which results in a disc spanning between $0.01-0.3 \, \rm au$.
Our composition tracking code then uses the unmodified relative abundances at a given radii of the re-scaled disc to initialize the bodies and to track the composition change from pebble accretion.

Differences in stellar evolution between G-type and M-dwarf stars will lead to different disc masses, lifetimes, mid-plane pressures, temperature profiles and thus different evolutionary tracks and timescales for dust condensation \citep{Ansdell2017, Siess2000, Picogna2021}.  However, extrapolating the relative abundances of the dust from a disc around a Sun-like star at a single epoch and re-scaling for disc size should be representative of the dust in an M-dwarf system with a solar composition, such as TRAPPIST-1 \citep{Gillon2016}, at some earlier epoch.  In the dust condensation code the dust condensates do not affect the subsequent evolution of the disc and to first approximation, the results reported here should be valid for the T1 system at some earlier time.  While the dust condensation is determined locally by density, mid-plane pressure and temperature, and differences in stellar evolution between G and M-type stars can lead to differences in these parameters, a fast evolution of the condensation only occurs when the disk local temperature is higher than the condensation temperature.  This can be seen in Fig. 7 of \cite{Li2020} and a clear condensation front can be seen over time.  As soon as the temperature reduces to below the condensation temperature, the abundance evolution becomes slowly varying. Thus, once the disc cools sufficiently the dust condensation is sensitive primarily to the initial disc composition.

Figure \ref{fig:profiles} compares the density of the gas disc $\rho$, temperature $T$, and mid-plane pressure $P$ profiles over 0.01--0.3\,au of the re-scaled profile around a Sun-like star from \cite{Li2020} and the analytic profiles used in the $n$-body simulations at $t=0$, as described in Section \ref{sec:models}.  
To derive the analytic $P$ profile we use

\begin{equation}
    P=\rho c_{\rm s}^2
\end{equation}
where $\rho = \Sigma/(H\sqrt{2 \pi})$ is the gas density, $H$ is the gas scale height and $c_{\rm s}$ is the sound speed.

The profiles used in \cite{Li2020} correspond to a disc shortly after disc formation and our disc profiles are for a more evolved disc and so differences can be seen between the profiles which may lead to differences in the dust evolution.  However, the density and pressure are still within approximately an order of magnitude of one another while the temperatures are below the condensation temperatures of the major elements thus, the relative abundances in condensed dust could remain similar in these two discs as both discs begin with a solar-like composition.

\bsp	
\label{lastpage}
\end{document}